%% file: DAC24-BiSMO.tex
\documentclass[sigconf,screen=true,anonymous=false,bookmarks=false,nonacm]{acmart}
\input{setting-acm}

\graphicspath{{./figs/}}
\setcopyright{none}

\usepackage{algpseudocodex}

\usepackage{expl3}
\makeatletter
\ams@newcommand{\multiint}[1]{\DOTSI\protect\MultiIntegral{#1}}
\renewcommand{\MultiIntegral}[1]{%
  \edef\ints@c{\noexpand\intop
    \ifnum#1=\z@\noexpand\intdots@\else\noexpand\intkern@\kern-0.3em\fi
    \replicate{#1-2}{\noexpand\intop\noexpand\intkern@\kern-0.3em}%
    \noexpand\intop
    \noexpand\ilimits@
  }%
  \futurelet\@let@token\ints@a
}
\makeatother
\ExplSyntaxOn
\cs_new:Npn \replicate #1 #2 { \prg_replicate:nn { #1 } { #2 } }
\ExplSyntaxOff

\usepackage{makecell}

\newcommand{\ie}{\textit{i}.\textit{e}., }

\begin{document}
\date{}

\title{
    Efficient Bilevel Source Mask Optimization
}

\author{Guojin Chen}
\affiliation{
    \institution{Chinese University of Hong Kong}
}
\email{gjchen21@cse.cuhk.edu.hk}

\author{Hongquan He}
\affiliation{
    \institution{Shanghai Tech. University}
}
\email{fallenhhq@gmail.com}

\author{Peng Xu}
\affiliation{
    \institution{Chinese University of Hong Kong}
}
\email{pxu22@cse.cuhk.edu.hk}

\author{Hao Geng}
\affiliation{
    \institution{Shanghai Tech. University}
}
\email{genghao@shanghaitech.edu.cn}

\author{Bei Yu}
\affiliation{
    \institution{Chinese University of Hong Kong}
}
\email{byu@cse.cuhk.edu.hk}

\input{doc/abstract}
\maketitle
\pagestyle{empty}

\input{doc/intro}
\input{doc/prelim}
\input{doc/algo}

\input{doc/exp}
\input{doc/conclu}

{
\bibliographystyle{IEEEtran-sim}
\bibliography{ref/Top-sim,ref/DL,ref/DFM,ref/NeRF,ref/bench,ref/CV,ref/SMO,ref/bilevel}
}

\end{document}

%% file: setting-acm.tex
\iftrue
\usepackage{titlesec}
\titlespacing\section{2pt}{5pt plus 1pt minus 1pt}{0pt plus 1pt minus 1pt}
\titlespacing\subsection{2pt}{5pt plus 1pt minus 1pt}{0pt plus 1pt minus 1pt}
\titlespacing\subsubsection{2pt}{5pt plus 1pt minus 1pt}{2pt plus 1pt minus 1pt}
\usepackage{enumitem}
\setlist{leftmargin=5.08mm}
\fi

\iftrue
\fi

\usepackage{lipsum}
\usepackage{graphicx}
\usepackage{amsmath}
\usepackage{footnote}
\usepackage{mathtools}
\usepackage{mathrsfs}     
\usepackage{comment}
\usepackage[subrefformat=parens,labelformat=parens]{subfig}
\usepackage{bm,bbm}
\usepackage{multirow}
\usepackage{threeparttable,booktabs}
\usepackage{blkarray}
\usepackage{tikz}
\usetikzlibrary{positioning,calc,fit,decorations.pathmorphing,shapes.geometric,shapes.gates.logic.US,calc}
\usepackage{tikz-3dplot}
\usepackage{balance}
\usepackage{courier}                                       
\usepackage[mathcal]{eucal}
\usepackage[]{algpseudocode}                               
\algrenewcommand\textproc{\texttt}
\makeatletter\let\float@addtolists\relax\makeatother
\usepackage{algorithm}
\renewcommand{\algorithmicrequire}{\textbf{Input:} }       
\renewcommand{\algorithmicensure} {\textbf{Output:}}       
\usepackage{cleveref}                                      
\usepackage{filecontents}                                  
\usepackage{pgfplots}
\usepackage{pgfplotstable}
\usepackage{pgf-pie}
\usepgfplotslibrary{groupplots}
\pgfplotsset{compat=newest}
\usepackage[figuresright]{rotating}
\usepackage{xcolor,colortbl}                               

\renewcommand{\vec}[1]{\boldsymbol{#1}}

\newcommand{\minisection}[1]{\vspace{.06in}\noindent{\textbf{#1}}}

\theoremstyle{plain}

\newtheorem{mylemma}{Lemma}

\theoremstyle{definition}
\newtheorem{mydefinition}{\textbf{Definition}}

\algrenewcommand\textproc{\texttt}

\usepackage[skip=1pt]{caption}            
\definecolor{CUHKorange}{RGB}{244,106,18} 
\definecolor{CUHKblue}{RGB}{0,111,190}    
\definecolor{CUHKgreen}{RGB}{0,127,128}   
\definecolor{CUHKred}{RGB}{228,46,36}     
\definecolor{CUHKyellow}{RGB}{198,148,34} 
\definecolor{CUHKdark}{RGB}{114,44,114}   
\definecolor{CUHKmiddle}{RGB}{144,44,144} 

\usepackage{tcolorbox}
\tcbuselibrary{skins,breakable}
    {\endtcolorbox}

%% file: doc/abstract.tex
\begin{abstract}
Resolution Enhancement Techniques (RETs) are critical to meet the demands of advanced technology nodes.
Among RETs, Source Mask Optimization (SMO) is pivotal, concurrently optimizing both the source and the mask to expand the process window.
Traditional SMO methods, however, are limited by sequential and alternating optimizations,
leading to extended runtimes without performance guarantees.
This paper introduces a unified SMO framework utilizing the accelerated Abbe forward imaging to enhance precision and efficiency.
Further, we propose the innovative \texttt{BiSMO} framework, which reformulates SMO through a bilevel optimization approach,
and present three gradient-based methods to tackle the challenges of bilevel SMO.
Our experimental results demonstrate that \texttt{BiSMO} achieves a remarkable 40\% reduction in error metrics
and 8$\times$ increase in runtime efficiency, signifying a major leap forward in SMO.
\end{abstract}

%% file: doc/intro.tex
\section{Introduction}
\label{sec:intro}
Lithography, vital for semiconductor manufacturing, advances integrated circuit (IC) development. The semiconductor industry's drive for miniaturization and efficiency challenges traditional lithography in creating finer patterns, emphasizing the need for resolution enhancement techniques (RETs) to meet advanced semiconductor requirements.
Among various RETs such as sub-resolution assist features (SRAFs)~\cite{OPC-USP2009-Wallace},
optical proximity correction (OPC)~\cite{OPC-DAC2014-Gao}, and source mask optimization (SMO),
SMO stands out due to its broader solution space.
SMO uniquely optimizes both illumination source and mask pattern,
ensuring pivotal lithographic fidelity vital for advancing next-generation IC manufacturing.

\begin{figure}[tbp]
  \centering
  \includegraphics[width=.99\linewidth]{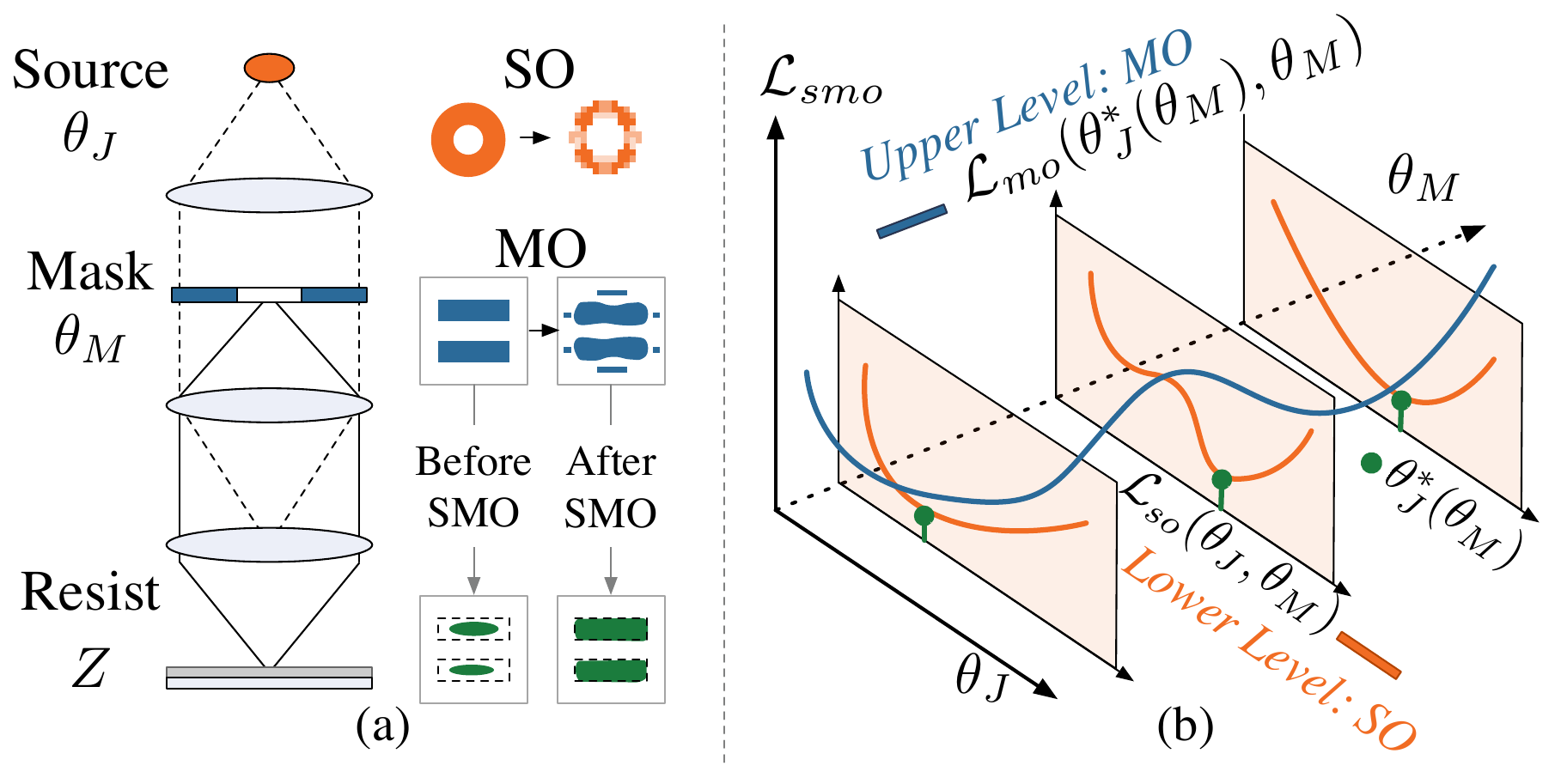}
  \caption{(a) The forward lithography and SMO process. (b) Bilevel SMO with upper-level MO and lower-level SO.}
  \label{fig:litho_bilevel}
\end{figure}

As shown in \Cref{fig:litho_bilevel}(a), SMO integrates source optimization (SO), mask optimization (MO), and their iterative optimization refinement.
The efficiency and performance of SMO are primarily influenced by two factors: the imaging model and the optimization strategy.
Central to both SO and MO are the forward imaging models: Abbe's~\cite{Fuehner2009abbe} and Hopkins'~\cite{cobb1995sum}, each with unique computational characteristics. Abbe's model, celebrated for its precision, demands extensive computation through the summation of intensities from discrete source points. Conversely, Hopkins' model, employing truncated singular value decomposition (SVD), reduces computational load but is unsuitable for SO due to its loss of source information.
Beyond isolated SO and MO, the strategy employed in their combined optimization significantly influences the optimization trajectory and solution space, thereby affecting SMO outcomes.

Within MO, Hopkins' model is foundational to various methods.
This includes MOSAIC~\cite{OPC-DAC2014-Gao} which blends design target and process window considerations, and techniques such as GAN-OPC~\cite{OPC-DAC2018-Yang}, DAMO~\cite{OPC-ICCAD2020-DAMO}, Neural-ILT~\cite{OPC-ICCAD2020-NeuralILT}, and DevelSet~\cite{OPC-ICCAD2021-DevelSet} that employ deep neural networks for enhanced performance.
Hardware acceleration strategies like those in GPU-LS~\cite{OPC-DATE21-Yu} and Multi-ILT~\cite{multiilt2023sun} leverage GPU and parallel acceleration to expedite Hopkins-based MO tasks.
To the best of our knowledge, the potential of Abbe-based MO optimization and acceleration remains unexplored.

For SO, the impracticality of Hopkins' model necessitates exclusive reliance on Abbe's.
Previous SO strategies have employed compressive sensing~\cite{wang2020csfasticc} and sampling-based methods~\cite{sampleICC2022sun} for complexity reduction.
However, the benefit is limited.
By contrast, MO using the Hopkins model with GPU or DNN acceleration achieve average optimization times of five seconds,
while SO dependent on Abbe's model typically exceed 30 minutes.
Due to its computational intensity, accelerating Abbe's model is essential.

In SMO, alongside the imaging model, the co-optimization strategy is a critical determinant of performance.
Since the inception of SMO in early 2010s, the alternating minimization (AM) strategy has been a dominant approach in SMO~\cite{wang2020csfasticc,ding2020gradient,sampleICC2022sun}.
As shown in \Cref{fig:traditional_bilevel_flow}(a), AM method involves isolated iterative minimization of source parameters across multiple SO epochs while maintaining constant mask variables, followed by mask optimization in MO epochs with the source fixed, repeating until convergence.
However, AM strategy's simplicity does not guarantee effectiveness.
Due to the concurrent impact of source and mask parameters on the aerial image,
AM's localized focus can result in SMO being confined to local minima, and the absence of global, gradient-based guidance can prolong convergence, necessitating more iterations.
These limitations highlight the need for more advanced co-optimization strategies in SMO.

To address SMO challenges,
we utilize the Abbe model for forward imaging,
leveraging its capability for concurrent SO and MO gradient computation,
and its superior precision due to avoidance of approximate decomposition.
We counterbalance the Abbe model's intensive computational demand with GPU acceleration.
This enables us to develop an Abbe-based unified SMO framework incorporating process window considerations,
consequently enhancing SMO outcomes, improving efficiency, and reducing process variability.
Furthermore, we innovatively reconceptualize the SMO problem as a bilevel optimization (BLO) challenge to gain a better co-optimization strategy,
as depicted in \Cref{fig:litho_bilevel}(b).
BLO, a hierarchical mathematical program, is defined as an optimization problem where the feasible region is constrained by another nested optimization problem.
It is widely applied in areas such as hyperparameter optimization~\cite{lorraine2020optimizing},
neural architecture search~\cite{zhang2021idarts},
and multitask or meta-learning~\cite{choe2023making}.
We then propose three novel gradient-based bilevel SMO solutions,
featuring global perspectives achieved through MO and SO gradient fusion.
These solutions demonstrate substantial improvements in efficiency and accuracy over the traditional SMO approaches.
Our primary contributions are as follows:
\begin{itemize}
  \item We establish the first unified Abbe-based SMO framework incorporating process window considerations, significantly accelerating Abbe imaging through parallel computation, achieving speeds comparable to Hopkins' method.
  \item We pioneer the modeling of SMO as a unified bilevel framework, developing three efficient gradient-based methods with global perspectives and improved exploration of solution space, superseding the conventional SMO.
  \item Our experimental results indicate that, compared to state-of-the-art (SOTA) SMO methods~\cite{sampleICC2022sun}, our approach reduces error metrics by approximately 40\% and achieves an eightfold increase in throughput.
  In comparison to SOTA MO methods~\cite{multiilt2023sun}, our error metrics are half as large.
\end{itemize}

%% file: doc/prelim.tex
\section{Preliminaries}
\label{sec:prelim}

\subsection{Lithography Simulation}
In optical lithography systems, the intensity of aerial image $I(x, y)$ on the wafer plane can be formulated via lithography theory~\cite{cobb1995sum} as:
\begin{equation}
  \label{eq:litho_imaging_full}
  \resizebox{.92\linewidth}{!}{$
  \begin{aligned}
  {I}(x, y)=&\!\multiint{6}_{-\infty}^{\infty}{J}(f, g) O(f^{\prime}, g^{\prime}) O^*(f^{\prime \prime}, g^{\prime \prime}){H}(f\!+\!f^{\prime}, g\!+\!g^{\prime})\\
  &{H}^*(f\!+\!f^{\prime\prime},\!g+g^{\prime \prime})\exp(-i2\pi((f^{\prime}\!-\!f^{\prime\prime})x+ (g^{\prime}\!-\!g^{\prime\prime})y))\\
  &\mathrm{d}f ~\mathrm{d}g ~\mathrm{d}f^{\prime} ~\mathrm{d}g^{\prime} ~\mathrm{d}f^{\prime\prime} ~\mathrm{d}g^{\prime\prime},
  \end{aligned}$}
\end{equation}
where $\vec{J}$ is the illumination source. $\vec{H}$ is projection system's optical transfer function.
$O(f^\prime, g^\prime)$ captures the binary mask pattern $M(x, y)$'s frequency spectrum,
derived via a $2D$ fast Fourier transform (\texttt{FFT}), $\mathcal{F}(\cdot)$.
$^*$ signifies the Hermitian transpose. $(x, y)$ denotes spatial coordinates, while $(f, g)$, $(f^{\prime}, g^{\prime})$, and $(f^{\prime \prime}, g^{\prime \prime})$ refer to the frequency coordinates of the source, mask spectrum, and its conjugate, respectively.
The formation of aerial image in \Cref{eq:litho_imaging_full} is computed using two distinct methods: Abbe's and Hopkins' method.

\minisection{Abbe's Approach:}
Abbe's approach, also known as the source points integration approach, discretizes the source space and independently computes the contribution of each source point, subsequently summing these contributions to form the aerial image.
Regardless of the discretization technique, the source can hence be represented as a set of source points $\{(f_\sigma, g_\sigma; j_\sigma)\}$,
where each source point defines a pair of spatial frequencies and its discrete magnitude $j_\sigma \in [0, 1]$.
By setting
$
\vec{A}_{(f, g)}(f^\prime, g^\prime) = {H}(f+f^{\prime}, g+g^{\prime}) O(f^{\prime}, g^{\prime}),
$
and applying the Inverse Fast Fourier Transform (\texttt{IFFT}),
the total intensity in \Cref{eq:litho_imaging_full} can be formulated in Abbe's approach:
\begin{equation}
  \begin{aligned}
  {I}(x, y) = \sum_{\sigma} j_\sigma \vert \vec{A}_{(f_\sigma,g_\sigma)}(x, y) \vert^2.
  \end{aligned}
  \label{eq:abbe_short}
\end{equation}
\minisection{Hopkins' Approach:}
Hopkins' approach separates the calculation of source and projection system from the processing of the mask for \Cref{eq:litho_imaging_full}.
It formulates the source and projector into the \textit{transmission cross-coefficients} ($TCC$),
as defined by:
\begin{equation}
  \resizebox{.9\linewidth}{!}{$
  \begin{aligned}
    {TCC}=\!\iint_{-\infty}^{\infty}{J}(f,g)
    {H}(f\!+\!f^{\prime}\!,g\!+\!g^{\prime}){H}^*(f\!+\!f^{\prime\prime},\!g\!+\!g^{\prime\prime})\mathrm{d}f\mathrm{d}g.
  \end{aligned}
  $}
  \label{eq:tcc_generation}
\end{equation}
The \textit{Sum of Coherent Systems} (SOCS)~\cite{cobb1995sum} provides an approximation to the Hopkins imaging equations, simplifying the ${TCC}$ spectrum using SVD.
Due to the rapid decay of eigenvalue $\kappa_q$ with $q$, only the top truncated $Q$ eigenvalues are retained.
By applying the \texttt{IFFT}, SOCS can be expressed in spatial domain as:
\begin{equation}
  {I}(x, y)=\sum_{q=1}^Q \kappa_q|{\phi}_q(x, y) \otimes {M}(x, y)|^2,
  \label{eq:socs_spatial}
\end{equation}
where ${\phi}_q(x, y)$ is the spatial distribution of eigenvector $\vec{\Phi_q}$. Here, $\otimes$ denotes convolution and $|\cdot|$ is the absolute operator.
\minisection{Hopkins' Approach vs. Abbe's Approach:}
\label{subsubsec:hopkins-vs-abbe}
In \Cref{eq:socs_spatial}, Hopkins' method reduces computational demands from Abbe's \(\mathcal{O}(N_j^2 \cdot N_m^4)\) to \(\mathcal{O}(Q \cdot N_m^4)\), with \(Q < N_j^2\) for source \(\vec{J} \in \mathbb{R}^{N_j \times N_j}\) and mask \(\vec{M} \in \mathbb{R}^{N_m \times N_m}\).
Under identical optical conditions, Hopkins' method outperforms Abbe's in speed, leading to its preference in various MO algorithms~\cite{OPC-DAC2014-Gao,OPC-DAC2018-Yang,OPC-ICCAD2020-DAMO,OPC-ICCAD2020-NeuralILT,OPC-ICCAD2021-DevelSet,OPC-DATE21-Yu,multiilt2023sun}.
However, Hopkins' reliance on truncated SVD, as shown in \Cref{eq:socs_spatial}, prevents SO due to the inability to calculate source gradients.
In contrast, Abbe's method (\Cref{eq:abbe_short}) inherently suits SO by summing the impacts of all source points to form the aerial image. Moreover, Abbe's richer source information enhances lithography precision, thus improving MO outcomes. Consequently, Abbe's method is essential for SO, higher MO precision, and indispensable for gradient-based bilevel SMO.

\subsection{Evaluation Metrics}

\begin{mydefinition}[Squared $L_2$ Error (L2)]
Given target pattern $\vec{Z}_t$ and resist image under nominal process condition $\vec{Z}$,
the squared $L_2$ error is calculated as $\Vert \vec{Z} - \vec{Z}_t \Vert_2^2$.
\end{mydefinition}

\begin{mydefinition}[Process Variation Band (PVB)]
PVB~\cite{OPC-ICCAD2013-Banerjee} is used in manufacturing to represent the expected range of variation in a production process.
PVB denotes the $\operatorname{XOR}$ area between the resist images $\vec{Z}_{\min}$ and $\vec{Z}_{\max}$ under the \textit{min} and \textit{max} process conditions.
\end{mydefinition}

\begin{mydefinition}[Edge Placement Error (EPE)]
EPE~\cite{OPC-ICCAD2013-Banerjee} refers to the deviation between the intended position of a feature on a wafer and its actual position after lithography.
\end{mydefinition}

%% file: doc/algo.tex
\section{Algorithm}
\label{sec:algo}
In this section, we detail the core contributions of our study: the development and acceleration of an Abbe-based SMO framework with process window considerations
in \Cref{subsec:abbe-based-smo-framework}.
Furthermore, in \Cref{subsec:bismo}, we offer a thorough derivation and comparison of gradient-based bilevel SMO,
where the upper-level MO gains a global perspective by gradient fusion, enabling more informed decision-makeing by solving the lower-level SO.
\subsection{Abbe-based Unified SMO Framework}
\label{subsec:abbe-based-smo-framework}
The Hopkins model, hindered by the need for frequent, inefficient, and non-differentiable SVD truncation of the TCC, limits effective gradient-based co-optimization of source and mask.
Addressing this, we introduce a unified SMO framework employing the Abbe model, which facilitates efficient, joint SMO without the TCC's burdensome processing.
The framework is further enhanced by incorporating parallel computing techniques for acceleration.

\minisection{Abbe-based unified SMO:}
Utilizing freeform illumination, the pixelated source point is denoted as \(J(f,g) \in [0, 1]\), while binary mask values \(M(x, y)\) are either 0 or 1.
To render the SMO framework differentiable, we introduce optimization parameters \(\vec{\theta}_J\) for Source \(\vec{J}\) and \(\vec{\theta}_M\) for Mask \(\vec{M}\), where \(\vec{\theta}_J \in \mathbb{R}^{N_j \times N_j}\), \(\vec{\theta}_M \in \mathbb{R}^{N_m \times N_m}\), and both parameters can assume any real value. Here, \(N_j\) and \(N_m\) represent the dimensions of the source and mask.
Appropriate activation and initialization enable deriving source and mask from these parameters.
The $\texttt{Sigmoid}$ function $\sigma(x) = 1 / \left(1 + \exp(-x)\right)$  is employed for both grayscale source and binary mask, as listed in \Cref{tab:activation_initialization}.
\begin{table}[!htp]\centering
  \caption{The activation and initialization for Abbe-imaging.}\label{tab:activation_initialization}
  \setlength{\tabcolsep}{3pt}
  \renewcommand{\arraystretch}{1.1}
  \resizebox{\linewidth}{!}{
  \begin{tabular}{|l|l|l|}\hline
  &Activation &Initialization \\\hline
  Mask $\vec{M}$   &$\vec{M} = \sigma(\alpha_m \cdot \vec{\theta}_M)$ & $\vec{\theta}_M(x, y) = {m_0}, \text{ if } M_0(x, y) = 1; \text{else } -\!m_0.$\\
  Source $\vec{J}$ &$\vec{J} = \sigma(\alpha_j \cdot \vec{\theta}_J)$ & $\vec{\theta}_J(f, g) = {j_0}, \text{  if  } J_0(f, g) = 1; \text{  else } -\!j_0.$ \\\hline
  \end{tabular}}
\end{table}

\noindent In \Cref{tab:activation_initialization}, \(\alpha_m\) and \(\alpha_j\) are the sigmoid steepness.
Initial values for \(\vec{\theta}_M\) are assigned as either $m_0$ or $-m_0$ based on initial mask pattern $\vec{M}_0$.
Typically, the initial mask pattern is the same as the binary target pattern \(\vec{Z}_t\).
This mask initialization also facilitates SRAF generation during MO.
Grayscale source \(\vec{J}\) requires careful selection of steepness \(\alpha_j\) and initialization $j_0$ to maintain its grayscale property.
Hyperparameters are detailed in \Cref{sec:exp}.
The shape of initial source pattern \(\vec{J}_0\) is derived from parametric templates like annular, quasar, or dipole, characterized by outer and inner radii \(\sigma_o\) and \(\sigma_i\).
Although the \texttt{Cosine} function is an alternative source activation, its use may lead to training instability due to gradient issues, leading us to prefer the \texttt{Sigmoid} function.
The transfer function $H(f, g)$ can be accurately characterized by a low-pass filter, expressed as:
\begin{equation}
  H(f, g)=\begin{cases}
    1, & \text{if } \sqrt{f^2+g^2} \leq \frac{{N\!A}}{\lambda}, \\
    0, & \text{otherwise},
  \end{cases}
  \label{eq:psf}
\end{equation}
where the cut-off frequency is determined by the projection system's numerical aperture ${N\!A}$ and the illumination wavelength $\lambda$.
By integrating \Cref{eq:abbe_short}, \Cref{eq:psf} and \Cref{tab:activation_initialization},
we formulate the Abbe forward imaging \(f_{abbe}\), determining the aerial image \(\vec{I}\) as a function of the optimization parameters for source $\vec{\theta}_J$ and mask $\vec{\theta}_M$:
$\vec{I} = f_{abbe}(\vec{\theta}_J, \vec{\theta}_M).$
Then we can utilize a straightforward threshold model for resist modeling.
The \texttt{Sigmoid} activation is also adopted to ensure a smooth transition and maintain differentiability:
\begin{equation}
  \vec{Z} = \sigma(\beta\cdot(\vec{I} - {I}_{tr})),
\end{equation}
where $\vec{Z}$ represents the resist pattern, ${I}_{tr}$ denotes the intensity threshold, and $\beta$ is the steepness.

We have now established a complete Abbe forward imaging model that maps the parameters \(\vec{\theta}_J\) and \(\vec{\theta}_M\)
to the aerial image \(\vec{I}\) and the resist image \(\vec{Z}\). To realize SMO, it is essential to define the corresponding objective function and optimization method.
We employ the mean squared loss to quantify the discrepancy between the resist pattern $\vec{Z}$ and the target pattern $\vec{Z}_t$:
\begin{equation}
  \mathcal{L}_2 = \Vert \vec{Z} - \vec{Z}_t \Vert^2.
\end{equation}
In alignment with \cite{OPC-DAC2014-Gao} and considering a \(\pm2\%\) dose range, we pioneer the integration of process window considerations into Abbe-based SMO to mitigate process variation via PVB loss.
By substituting $\vec{M}_{\min}\!=\!d_{\min}\cdot\sigma(\alpha_m\cdot\vec{\theta}_M)$ and $\vec{M}_{\max}\!=\!d_{\max}\cdot\sigma(\alpha_m\cdot\vec{\theta}_M)$ into $f_{abbe}$,
we obtain the resist patterns $\vec{Z}_{\min}$, $\vec{Z}_{\max}$ under minimum $d_{\min}$ and maximum $d_{\max}$ process conditions.
The PVB loss is formulated as:
\begin{equation}
  \mathcal{L}_{pvb} = \Vert \vec{Z}_{\max} - \vec{Z}_t \Vert^2 + \Vert \vec{Z}_{\min} - \vec{Z}_t\Vert^2.
\end{equation}
Consequently, the comprehensive SMO loss $\mathcal{L}_{smo}$ is formulated as:
\begin{equation}
  \mathcal{L}_{smo} \coloneqq \mathcal{L}_{so} \coloneqq \mathcal{L}_{mo} = \gamma \mathcal{L}_2 + \eta \mathcal{L}_{pvb},
\end{equation}
where $\gamma$ and $\eta$ are weighting factors for the respective loss components.
SO loss $\mathcal{L}_{so}$ and MO loss $\mathcal{L}_{mo}$ can utilize same objective functions.
The SMO problem is thus defined as:
\begin{equation}
  (\hat{\vec{\theta}_J}, \hat{\vec{\theta}_M}) = \underset{(\vec{\theta}_J, \vec{\theta}_M)}{\operatorname{argmin}}~ \mathcal{L}_{smo}(\vec{\theta}_J, \vec{\theta}_M),
  \label{eq:smo_problem_formulation}
\end{equation}
where $\hat{\vec{\theta}_J}, \hat{\vec{\theta}_M}$ represent the optimal parameter values for the source and mask, respectively.

\minisection{Abbe acceleration.}
\label{subsubsec:abbe_acceleration}
The primary computational bottleneck in SMO is the forward imaging model and its gradient calculations.
As noted in \Cref{eq:abbe_short,eq:socs_spatial}, contributions from source points can be independently calculated, making the complexity ratio between Abbe's and Hopkins's models \(\sigma / Q\), where \(\sigma \in [0, N_j^2]\) represents the number of effective source points (\ie where $j_\sigma > 0$).
Both models can be accelerated with parallel computing, using multicore CPUs or GPUs.
Theoretically, the parallel computation time ratio is \(\lceil  \frac{\sigma}{P} \rceil / \lceil \frac{Q}{P} \rceil\), where \(P\) denotes the maximum number of parallel threads and \(\lceil \cdot \rceil\) is the ceiling operator.
This suggests that Abbe's runtime can match Hopkins' if \(P \geq \sigma\). In our implementation, GPUs are utilized for parallel computation of each effective source point's contribution to the aerial image, owing to their greater thread parallelism, larger memory bandwidth, and faster \texttt{FFT} or \texttt{IFFT} operations compared to multicore CPUs. Experimental results in \Cref{subsubsec:runtime_comparison} demonstrate that our Abbe-imaging model achieves runtime performance comparable to Hopkins' model.

\subsection{Efficient Bilevel SMO}
\label{subsec:bismo}

\begin{figure}[tbp]
  \centering
  \includegraphics[width=.99\linewidth]{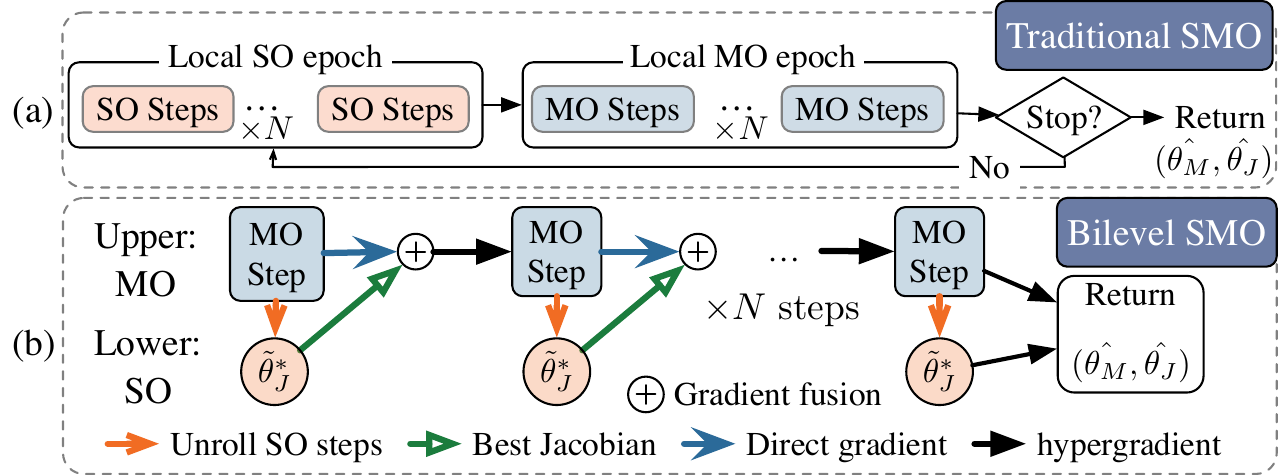}
  \caption{(a)Previous AM-SMO flow. (b)Our BiSMO flow.}
  \label{fig:traditional_bilevel_flow}
\end{figure}

\minisection{Previous alternating minimization-based SMO:}
Since the introduction of SMO technology, the significant computational demands have compelled previous methods to compromise on a simple alternating minimization (AM) strategy.
As illustrated in \Cref{alg:alternating-SMO}, AM-based SMO (\texttt{AM-SMO}) alternates between two minimization cycles, updating parameters \(\vec{\theta}_J\) and \(\vec{\theta}_M\) sequentially. This process iterates until reaching the specified convergence criteria for SO and MO, as shown in \Cref{fig:traditional_bilevel_flow}(a).
However, \texttt{AM-SMO} has several notable drawbacks:
1) \texttt{AM-SMO} tends to converge to local minima due to its narrow focus on localized aspects of SO or MO, ignoring the global structure of the problem.
2) The convergence is often slow because the source and mask are highly interdependent. Adjusting \((\vec{\theta}_M)_k\) as per \cref{alg1:line:mo-iters-with-source-fixed} makes \((\vec{\theta}_J)_k\) suboptimal, requiring numerous iterations for stabilization.
3) The absence of global gradient guidance complicates establishing effective early stopping criteria, often resulting in either prolonged optimization or suboptimal convergence.
\begin{algorithm}[h]
  \caption{\small Alternating Minimization-based SMO (\texttt{AM-SMO})~\cite{wang2020csfasticc,sampleICC2022sun}}
  \begin{algorithmic}[1]
      \For{$k = 1, 2, 3, \ldots$;} \Comment{Alternating SO \& MO.}
          \While{{not converged}} \Comment{SO iterations.}
            \State $(\vec{\theta}_J)_k \gets {\operatorname{argmin}}_{\vec{\theta}_J} \mathcal{L}_{so}\left(\vec{\theta}_J, (\vec{\theta}_M)_{k - 1}\right)$; \Comment{$\vec{\theta}_M$ is fixed.}
          \EndWhile
          \While{{not converged}} \Comment{MO iterations.}
            \State $(\vec{\theta}_M)_{k} \gets \operatorname{argmin}_{\vec{\theta}_M} \mathcal{L}_{mo}\left((\vec{\theta}_J)_k, \vec{\theta}_M\right)$;  \Comment{$\vec{\theta}_J$ is fixed.} \label{alg1:line:mo-iters-with-source-fixed}
          \EndWhile
      \EndFor
      \Return $(\vec{\theta}_J)_k, (\vec{\theta}_M)_k$.
  \end{algorithmic}
  \label{alg:alternating-SMO}
\end{algorithm}

\minisection{Proposed bilevel SMO:}
\Cref{eq:smo_problem_formulation} frames SMO as a typical multivariate optimization problem.
Yet, the \texttt{AM-SMO}, as detailed in \Cref{alg:alternating-SMO}, often leads to suboptimal results due to its localized focus.
A gradient-based approach with a global perspective is essential to overcome these limitations.
In SMO, SO and MO have a nested relationship.
The goal is to rapidly and effectively determine the optimal source response for each altered mask,
offering MO global gradient direction to boost overall SMO efficiency.
This scenario is particularly well-suited for a bilevel optimization approach.
From a bilevel viewpoint, the upper-level MO, constrained by the optimal solutions from the lower-level SO, forms a dependent hierarchy, as illustrated in \Cref{fig:litho_bilevel}(b) and \Cref{fig:traditional_bilevel_flow}(b).
This structure allows the MO to offer a global perspective by solving SO, guiding the optimization beyond local minima.
Consequently, \Cref{eq:smo_problem_formulation} can be reformulated in a bilevel context:
\begin{flalign}
  \begin{aligned}
    \min_{\vec{\theta}_M} &\ \mathcal{L}_{mo}(\vec{\theta}_J^{*}(\vec{\theta}_M), \vec{\theta}_M),  &\Comment{Upper-Level: MO} \\
    \textup{s.t.}     & \ \vec{\theta}_J^{*}(\vec{\theta}_M) = {\operatorname{argmin}}_{\vec{\theta}_J}~\mathcal{L}_{so}(\vec{\theta}_J, \vec{\theta}_M). &\Comment{Lower-Level: SO}
  \end{aligned}
  \label{eq:bilevel-smo-defination}
\end{flalign}
In BLO, the inner and outer loops are termed lower-level and upper-level subproblems.
Here the inner loop seeks optimal source parameters $\vec{\theta}_J^{*}$ for the current $\vec{\theta}_M$
while the outer loop endeavors to optimize the mask parameters $\vec{\theta}_M$ with the best-response source $\vec{\theta}_J^*(\vec{\theta}_M)$.
The gradient of the outer loop in bilevel SMO, also referred to as the hypergradient,
which is derived from the fusion of gradients from the upper and lower levels,
is then calculated as:
\begin{equation}
  \nabla_{\vec{\theta}_M}\mathcal{L}_{mo}=\frac{\partial \mathcal{L}_{mo}}{\partial \vec{\theta}_M}+\frac{\partial \mathcal{L}_{mo}}{\partial \vec{\theta}_J}\frac{\partial \vec{\theta}_J^{*}(\vec{\theta}_M)}{\partial \vec{\theta}_M}.
  \label{eq:bileve_hypergradient}
\end{equation}
Within the Abbe-based SMO framework introduced in \Cref{subsec:abbe-based-smo-framework},
the direct gradient $\frac{\partial \mathcal{L}{mo}}{\partial \vec{\theta}_M}$ and $\frac{\partial \mathcal{L}{mo}}{\partial \vec{\theta}_J}$ can be efficiently computed.
However, two principal challenges arise: (1) the precise approximation of the SO optimal solution $\vec{\theta}_J^{*}(\vec{\theta}_M)$,
and (2) differentiating the best-response Jacobian: $\frac{\partial \vec{\theta}_J^{*}(\vec{\theta}_M)}{\partial \vec{\theta}_M}$.
To address the former, we approximate \(\vec{\theta}_J^*\) by unrolling a few SO gradient steps \(T\), significantly reducing the computational cost.
Fortunately, extensive research~\cite{lorraine2020optimizing,zhang2021idarts,choe2023making} indicates that BLO, with weight sharing, can effectively adapt \(\vec{\theta}_J\) to \(\vec{\theta}_J^*\) with a small unrolling step \(T\).
For the latter issue, we propose three methods to compute the best-response Jacobian:
bilevel SMO using finite difference (\texttt{BiSMO-FD}),
\texttt{BiSMO-NMN} utilizing Neumann series, and \texttt{BiSMO-CG} using conjugate gradients.

\subsubsection{BiSMO-FD:}
\label{subsubsec:bismo-fd}
The finite difference (\texttt{FD}) strategy uses a single inner SO step,
$\vec{\theta}_J^{*}(\vec{\theta}_M) = \vec{\theta}_J - \xi\nabla_{\vec{\theta}_J}\mathcal{L}_{so}$, and obtaining $\frac{\partial\vec{\theta}_J^{*}(\vec{\theta}_M)}{\partial\vec{\theta}_M} = -\xi\frac{\partial^2\mathcal{L}_{so}}{\partial\vec{\theta}_M\partial\vec{\theta}_J}$,
the \texttt{BiSMO-FD} calculates the hypergradient as:
\begin{equation}
  \nabla_{\vec{\theta}_M}\mathcal{L}_{mo}^{\texttt{FD}}=\frac{\partial \mathcal{L}_{mo}}{\partial \vec{\theta}_M}-\xi\frac{\partial \mathcal{L}_{mo}}{\partial \vec{\theta}_J}\frac{\partial^2 \mathcal{L}_{so}}{\partial \vec{\theta}_M\partial \vec{\theta}_J},
  \label{eq:smo_darts_hypergradient}
\end{equation}
where $\xi$ is the inner-loop learning rate for $\vec{\theta}_J$.
While increasing the number of inner SO steps can lead to a more precise approximation for $\vec{\theta}_J^*$,
akin to \texttt{AM-SMO}, it results in a linear increase in memory and computational load, becoming impractical due to the need to store optimization paths and all intermediate gradients for differentiation following the chain rule.

In contrast, using the implicit function theorem (IFT), the hypergradient can be computed without retaining intermediate gradients, thus independent of the optimization path and significantly reducing memory usage.
The IFT-based hypergradient for SMO can be formulated as the following lemma.
\begin{mylemma}{Implicit Function Theorem:}
Consider $\vec{\theta}_J^{*}(\vec{\theta}_M)$ defined in \Cref{eq:bilevel-smo-defination}, with first-order optimality condition $\frac{\partial\mathcal{L}_{so}(\vec{\theta}_J^{*},\vec{\theta}_M)}{\partial \vec{\theta}_J}=0$,
\resizebox{\linewidth}{!}{$
\begin{aligned}
&\frac{\partial}{\partial \vec{\theta}_M}\!\left[\frac{\partial\mathcal{L}_{so}(\vec{\theta}_J^{*}(\vec{\theta}_M), \vec{\theta}_M)}{\partial \vec{\theta}_J} \right]\!=\!0,\ \Rightarrow
\frac{\partial^2 \mathcal{L}_{so}}{\partial \vec{\theta}_M\partial \vec{\theta}_J}\!+\!\frac{\partial^2 \mathcal{L}_{so}}{\partial \vec{\theta}_J\partial \vec{\theta}_J}\frac{\partial (\vec{\theta}_J^{*}(\vec{\theta}_M))}{\partial\vec{\theta}_M}=0,\\
&\Rightarrow\text{best-response Jacobian: } \frac{\partial (\vec{\theta}_J^{*}(\vec{\theta}_M))}{\partial\vec{\theta}_M}=-\left[ \frac{\partial^2 \mathcal{L}_{so}}{\partial\vec{\theta}_J\partial\vec{\theta}_J} \right]^{-1}\frac{\partial^2 \mathcal{L}_{so}}{\partial\vec{\theta}_M \partial\vec{\theta}_J}.
\end{aligned}
$}
With \Cref{eq:bileve_hypergradient}, we have hypergradient formulated by:
\begin{equation}
    \nabla_{\vec{\theta}_M}\mathcal{L}_{mo}=\frac{\partial \mathcal{L}_{mo}}{\partial \vec{\theta}_M}-\frac{\partial \mathcal{L}_{mo}}{\partial \vec{\theta}_J}\left [ \frac{\partial^2 \mathcal{L}_{so}}{\partial \vec{\theta}_J\partial \vec{\theta}_J} \right ]^{-1}\frac{\partial^2 \mathcal{L}_{so}}{\partial \vec{\theta}_M \partial \vec{\theta}_J}.
    \label{eq:ift_hypergradient}
  \end{equation}
\end{mylemma}
\noindent However, the inverse Hessian $\left [ \frac{\partial^2 \mathcal{L}_{so}}{\partial \vec{\theta}_J\partial \vec{\theta}_J} \right ]^{-1}$ in \Cref{eq:ift_hypergradient} is hard to calculate.
In \Cref{eq:smo_darts_hypergradient}, \texttt{BiSMO-FD} employs finite difference to naively approximate the inverse $\left [ \frac{\partial^2 \mathcal{L}_{so}}{\partial \vec{\theta}_J\partial \vec{\theta}_J} \right ]^{-1} = \xi \vec{\mathcal{I}}$, where $\vec{\mathcal{I}}$ denotes the identity matrix.
For a more precise approximation of the inverse, we introduce two IFT-based methods: Neumann series (\texttt{BiSMO-NMN}) and conjugate gradient (\texttt{BiSMO-CG}), to reformulate the hypergradient.
\subsubsection{BiSMO-NMN}
\label{subsubsec:bismo-nmn}
\begin{mylemma}
  Neumann series~\cite{lorraine2020optimizing}: With a matrix $\vec{A}$ that $\left \| \vec{\mathcal{I}}-\vec{A} \right \|<1$, we have $\vec{A}^{-1}=\sum_{k=0}^{\infty}(\vec{\mathcal{I}}-\vec{A})^k$.
  \label{lemma:neumannapprox}
\end{mylemma}
\noindent Based on \Cref{lemma:neumannapprox}, with small enough learning rate,
the hypergradient in \Cref{eq:ift_hypergradient} for \texttt{BiSMO-NMN} is formulated by:
  \begin{equation}
    \begin{aligned}
      \nabla_{\vec{\theta}_M}\mathcal{L}_{mo}
      & =\frac{\partial \mathcal{L}_{mo}}{\partial \vec{\theta}_M}-\frac{\partial \mathcal{L}_{mo}}{\partial \vec{\theta}_J}\sum_{k=0}^{\infty}\left [ \vec{\mathcal{I}}- \frac{\partial^2 \mathcal{L}_{so}}{\partial \vec{\theta}_J \partial \vec{\theta}_J} \right ]^k \frac{\partial^2 \mathcal{L}_{so}}{\partial \vec{\theta}_M \partial \vec{\theta}_J}. \\
    \end{aligned}
    \label{eq:smo_hyper_neuman}
  \end{equation}
\noindent The approximation of the hypergradient \(\nabla_{\vec{\theta}_M}\tilde{\mathcal{L}}_{mo}\) can be derived by considering only the first \(K\) terms of the Neumann series, thereby avoiding the need to calculate the inverse of the Hessian as:
\begin{equation}
  \nabla_{\vec{\theta}_M}\tilde{\mathcal{L}}^{\texttt{NMN}}_{mo}=\frac{\partial \mathcal{L}_{mo}}{\partial \vec{\theta}_M}- \frac{\partial \mathcal{L}_{mo}}{\partial \vec{\theta}_J}\sum_{k=0}^{K}\left [ \vec{\mathcal{I}} - \frac{\partial^2 \mathcal{L}_{so}}{\partial \vec{\theta}_J \partial \vec{\theta}_J} \right ]^k \frac{\partial^2 \mathcal{L}_{so}}{\partial \vec{\theta}_M \partial \vec{\theta}_J}.
  \label{eq:neumann_approx_hypergradient}
\end{equation}

\subsubsection{BiSMO-CG:}
Instead of calculating the Neumann series, another efficient way to approximate the inverse Hessian is to solve the linear systems.
Specifically, $\frac{\partial \mathcal{L}_{mo}}{\partial \vec{\theta}_J}\left [ \frac{\partial^2 \mathcal{L}_{so}}{\partial \vec{\theta}_J\partial \vec{\theta}_J} \right ]^{-1}$
can be computed as the solution to the linear system
$
\left [ \frac{\partial^2 \mathcal{L}_{so}}{\partial \vec{\theta}_J\partial \vec{\theta}_J} \right ]\vec{w} = \frac{\partial \mathcal{L}_{mo}}{\partial \vec{\theta}_J}.
$
The vector $\vec{w}$ can be obtained by solving the optimization problem:
\begin{equation}
  \min_{\vec{w}}\vec{w}^{\top}\left [ \frac{\partial^2 \mathcal{L}_{so}}{\partial \vec{\theta}_J\partial \vec{\theta}_J} \right ]\vec{w} - \vec{w}^\top\frac{\partial \mathcal{L}_{mo}}{\partial \vec{\theta}_J}.
\end{equation}
The conjugate gradient (\texttt{CG}) algorithm is well-suited for this task, given its efficient iteration complexity
and use of Hessian-vector products (\texttt{HVP}) for $\left [ \frac{\partial^2 \mathcal{L}_{so}}{\partial \vec{\theta}_J\partial \vec{\theta}_J} \right ]\vec{w}$.
Such \texttt{HVP} can be obtained cheaply without explicitly forming or storing the Hessian.
The hypergradient in \Cref{eq:ift_hypergradient} for \texttt{BiSMO-CG} is then computed as:
\begin{equation}
  \label{eq:hypergradient_cg}
  \resizebox{.9\linewidth}{!}{$
  \nabla_{\vec{\theta}_M}\tilde{\mathcal{L}}^{\texttt{CG}}_{mo}\!=\!\frac{\partial \mathcal{L}_{mo}}{\partial \vec{\theta}_M}\!-\!\left[\underset{\vec{w}}{\operatorname{argmin}}\left(\vec{w}^\top\frac{\partial^2\mathcal{L}_{so}}{\partial\vec{\theta}_J\partial \vec{\theta}_J}\vec{w}\!-\!\vec{w}^\top\frac{\partial\mathcal{L}_{mo}}{\partial\vec{\theta}_J}\right) \right]\frac{\partial^2\mathcal{L}_{so}}{\partial\vec{\theta}_M\partial\vec{\theta}_J}.
  $}
\end{equation}
\subsubsection{\texttt{BiSMO-FD} vs. \texttt{BiSMO-NMN} vs. \texttt{BiSMO-CG} vs. \texttt{AM-SMO}:}
\label{subsubsec:comparison-of-three-strategies}
The optimization flow of \texttt{BiSMO} is demonstrated in \Cref{fig:traditional_bilevel_flow}(b) and \Cref{alg:bismo}.
When $K = 0$, the $\nabla_{\vec{\theta}_M}\tilde{\mathcal{L}}^{\texttt{NMN}}_{mo}$ in \Cref{eq:neumann_approx_hypergradient} reduces to match $\nabla_{\vec{\theta}_M}\mathcal{L}_{mo}^{\texttt{FD}}$ in \Cref{eq:smo_darts_hypergradient}.
In \Cref{alg:bismo}, set $K=0; T=1;$ the \texttt{FD} can be executed through the same process as \texttt{NMN}.
Both \texttt{NMN} and \texttt{CG} use $T$ unroll SO steps to approximate $\vec{\theta}_J^{*}$ (\cref{alg2:line:unroll-so}), and employ \texttt{HVP} or Jacobian-vector product (\texttt{JVP}) for computational acceleration.
The key difference between them lies in approximating the inverse Hessian:
\texttt{NMN} uses the first \(K\) terms of the Neumann series (\cref{alg2:line:approximate-nmn}),
while \texttt{CG} applies \(K\) CG steps (\cref{alg2:line-approximate-cg}).
Compared to \texttt{AM-SMO}, the unroll strategy in \cref{alg2:line:unroll-so}, due to the small T, substantially reduces runtime by avoiding full SO cycle convergence.
Furthermore, in \cref{alg2:line:neumann_approx_hypergradient} and \cref{alg2:line:hypergradient_cg}, IFT-based gradient fusion provides \texttt{BiSMO} with a global perspective and a more thorough exploration of the solution space, thereby facilitating enhanced and accelerated convergence.

\begin{algorithm}[t]
    \caption{Bilevel SMO via \texttt{BiSMO-NMN}, \texttt{BiSMO-CG}}
    \label{alg:bismo}
    \small
    \begin{algorithmic}[1]
      \Require Unroll step $T$, stepsizes $\xi_{J}, \xi_{M}$, initializations $\vec{\theta}_J^0, \vec{\theta}_M^0,\vec{w}_0$, term $K$
      \Ensure $\vec{\theta}_J, \vec{\theta}_M$
      \While{not converged}
        \For{Inner step $t=1, \ldots ,T$} \Comment{Unroll $T$ steps of inner-SO.} \label{alg2:line:unroll-so}
          \State Update $\vec{\theta}_J^t \gets \vec{\theta}_J^{t-1}-\xi_J\nabla_{\vec{\theta}_J} \mathcal{L}_{so}(\vec{\theta}_J^{t-1}, \vec{\theta}_M)$; // Or Adam.
        \EndFor
        \State Approximate $\vec{\theta}_J^{*} \gets \vec{\theta}_J^t$;  Re-initialize $\vec{\theta}_J^0 \gets \vec{\theta}_J^{t}$;
      \If{\texttt{BiSMO-NMN}} \Comment{Hypergradient via \texttt{BiSMO-NMN}.}
      \State 1) Get $K$ Neumann series via \texttt{HVP}: $\frac{\partial \mathcal{L}_{mo}}{\partial \vec{\theta}_J}\sum_{k=0}^{K}\left [ \vec{\mathcal{I}} - \frac{\partial^2 \mathcal{L}_{so}}{\partial \vec{\theta}_J \partial \vec{\theta}_J} \right ]^k$; \label{alg2:line:approximate-nmn}
      \State 2) Get Jacobian-vector product \texttt{JVP} in \Cref{eq:neumann_approx_hypergradient};
      \State 3) $\nabla_{\vec{\theta}_M}\tilde{\mathcal{L}}^{\texttt{NMN}}_{mo} \gets$ \Cref{eq:neumann_approx_hypergradient}; \label{alg2:line:neumann_approx_hypergradient}
      \EndIf
      \If{\texttt{BiSMO-CG}} \Comment{Hypergradient via \texttt{BiSMO-CG}.}
      \State{1) Solve $\vec{w}_{K}$ from $\left [ \frac{\partial^2 \mathcal{L}_{so}}{\partial \vec{\theta}_J\partial \vec{\theta}_J} \right ]\vec{w} = \frac{\partial \mathcal{L}_{mo}}{\partial \vec{\theta}_J},$ via $K$ steps of CG starting from $\vec{w}_{0}$; then re-initialize $\vec{w}_0 \gets \vec{w}_K$}; \label{alg2:line-approximate-cg}
      \State{2) Get Jacobian-vector product \texttt{JVP}:$\left[\frac{\partial^2 \mathcal{L}_{so}}{\partial \vec{\theta}_M \partial \vec{\theta}_J}\right]\vec{w}$;}
      \State{3) $\nabla_{\vec{\theta}_M}\tilde{\mathcal{L}}^{\texttt{CG}}_{mo} \gets$ \Cref{eq:hypergradient_cg}}; \label{alg2:line:hypergradient_cg}
      \EndIf
      \State Update $\vec{\theta}_M \gets \vec{\theta}_M - \xi_{M}\nabla_{\vec{\theta}_M}\tilde{\mathcal{L}}_{mo}$; // Or Adam.
      \EndWhile
  \end{algorithmic}
\end{algorithm}

%% file: doc/exp.tex
\section{Experiments}
\label{sec:exp}
The \texttt{BiSMO} is implemented in \texttt{PyTorch} framework and tested on an Nvidia RTX4090 GPU card across three datasets as listed in \Cref{tab:data}.
The hyperparameters settings are as follows $\gamma\!=\!1000; \eta\!=\!3000; \lambda\!=\!193nm; N\!A\!=\!1.35; \sigma_o\!=\!0.95; \sigma_i\!=\!0.63. Q\!=\!24; N_j\!=\!35; N_m \!=\! 2048, \alpha_m \!=\! 9; m_0 \!=\! 1;  \alpha_j \!=\! 2; j_0 \!=\! 5; \beta \!=\! 30; P\!=\!256; \xi\!=\!\xi_{M}\!=\!\xi_{J}\!=\!0.1; K\!=\!5; T\!=\!3$.
All the tiles are converted to $2048 \times 2048$-pixel images.
\input{doc/res/dataset}

\subsection{Results Comparison with SOTA}

\input{doc/res/all_data}
\input{doc/res/epe_runtime}

We have conducted a comparative analysis of the performance between our Abbe-based MO and the previous SOTA MO methods DAC23-MILT~\cite{multiilt2023sun} and NILT~\cite{OPC-ICCAD2020-NeuralILT} in \Cref{tab:all_data} and \Cref{tab:epe_tat}.
Furthermore, we compare the performance of \texttt{BiSMO} with the previous SOTA \texttt{AM-SMO}~\cite{sampleICC2022sun,ding2020gradient}.
\texttt{AM-SMO} is implemented in two ways: one involves hybrid Abbe-SO and Hopkins-MO~\cite{ding2020gradient}, while the other employs the Abbe model for both SO and MO~\cite{sampleICC2022sun}.

\begin{figure}[htbp]
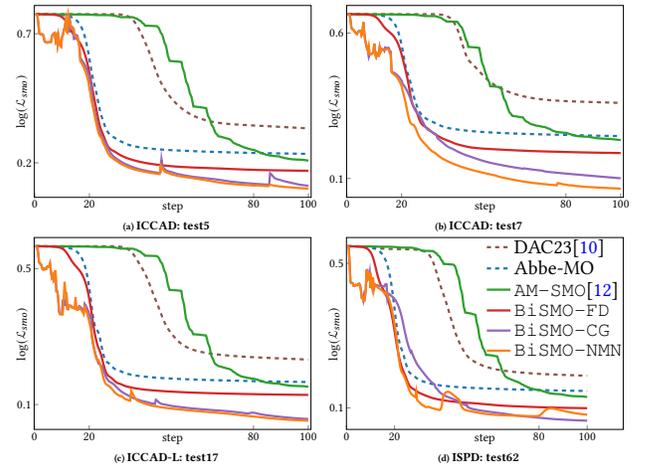

  \centering
  \resizebox{\linewidth}{!}{
    \renewcommand{\arraystretch}{0.5}
    \begin{tabular}{c}
      \subfloat[][{\normalsize ICCAD: test5}]{\input{figs/pgfplot/ICCAD_test5/ICCAD_test5}}
      \subfloat[][{\normalsize ICCAD: test7}]{\input{figs/pgfplot/ICCAD_test7/ICCAD_test7}} \\[-10pt]
      \subfloat[][{\normalsize ICCAD-L: test17}]{\input{figs/pgfplot/ICCAD-L_test17/ICCAD-L_test17}}
      \subfloat[][{\normalsize ISPD: test62}]{\input{figs/pgfplot/ISPD_test12/ISPD_test12} \label{fig:ispd-test12-cg-better}}
  \end{tabular}
  }
  \caption{Loss comparison between different MO methods (dashed lines) and SMO methods (solid lines).}
  \label{fig:loss_curves}
\end{figure}

\begin{figure}[tbp]
  \centering
  \includegraphics[width=.99\linewidth]{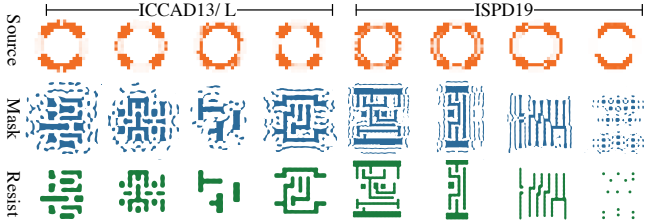}
  \caption{Result samples from ICCAD13 and ISPD19 datasets.}
  \label{fig:result_samples}
\end{figure}

To highlight advantages of \texttt{BiSMO}, \Cref{fig:loss_curves} shows log-scaled loss $\log(\mathcal{L}_{smo})$ convergence compared to SOTA MO~\cite{multiilt2023sun} and \texttt{AM-SMO}~\cite{sampleICC2022sun} using random cases from test datasets in \Cref{tab:data}, with a 0.01 learning rate.
Result samples are depicted in \Cref{fig:result_samples}, appropriately scaled and cropped to enhance visualization.

\minisection{Effectiveness of Abbe-MO:}
\Cref{tab:all_data} and \Cref{tab:epe_tat} demonstrate that our Abbe-MO achieved a 25\% reduction in L2, 19\% in PVB, and 24\% in EPE when compared to the SOTA MO DAC23-MILT~\cite{multiilt2023sun}.
Furthermore, within the \texttt{AM-SMO}, the Abbe-based SMO\cite{sampleICC2022sun} surpassed the Abbe-Hopkins hybrid SMO\cite{ding2020gradient} by decreasing L2 by 28\%, PVB by 21\%, and EPE by 29\%.
In \Cref{fig:loss_curves}, all cases indicate that Abbe-MO converges more rapidly and effectively than SOTA MO~\cite{multiilt2023sun}.
This can be attributed to the fact that truncated decomposition in Hopkins' approach leads to a loss of accuracy in lithography, thereby allowing our lossless Abbe method to achieve superior MO and SMO results.

\minisection{BiSMO vs. {AM-SMO} vs. MO:}
\Cref{tab:all_data}, \Cref{tab:epe_tat} and \Cref{fig:loss_curves} reveal that \texttt{BiSMO} variants significantly outperform \texttt{AM-SMO}\cite{ding2020gradient,sampleICC2022sun} in error reduction,
with \texttt{BiSMO-NMN} achieving decreases of 41\% in L2, 46\% in PVB, and 37\% in EPE,
and even the basic \texttt{BiSMO-FD} showing reductions of 36\% in L2, 34\% in PVB, and 27\% in EPE.
\Cref{fig:loss_curves} illustrates \texttt{AM-SMO}'s~\cite{sampleICC2022sun} `zigzag' loss curve, a result of its alternating optimization, which ultimately settles below Abbe-MO yet above \texttt{BiSMO} variants. This suggests that while \texttt{AM-SMO}'s broader solution space improves outcomes compared to pure MO.
However, its alternating approach risks entrapment in local minima, hindering it from reaching the optimal results achievable by \texttt{BiSMO}. This fact underscores the superior performance of the \texttt{BiSMO} method. Additionally, \texttt{BiSMO} demonstrates a significant improvement over the SOTA MO~\cite{multiilt2023sun}, achieving a $\sim$50\% reduction in all error metrics.

\minisection{Runtime comparison:}
\label{subsubsec:runtime_comparison}
In our implementation, Abbe-MO and Hopkins-MO have been accelerated to average 0.16s and 0.12s per MO iteration, aligning with theoretical derivations in \Cref{subsubsec:abbe_acceleration}.
As shown in \Cref{tab:epe_tat}, for parity, we have applied GPU acceleration to \texttt{AM-SMO}~\cite{ding2020gradient,sampleICC2022sun} with settings identical to \texttt{BiSMO}.
In their original implementations, the runtime for \cite{sampleICC2022sun} was 910s, and \cite{ding2020gradient} was 69 minutes.
Utilizing our accelerated Abbe imaging, this has been accelerated to 122.5s and 287s, respectively.
Despite these improvements, \texttt{BiSMO}, leveraging hypergradient, still achieves faster convergence than \cite{ding2020gradient,sampleICC2022sun},
boosting throughput by 8.3 times compared to the Abbe-based \texttt{AM-SMO}~\cite{sampleICC2022sun}.
Furthermore, it's about 19.5 times quicker than the Abbe-Hopkins hybrid \texttt{AM-SMO}~\cite{ding2020gradient}, which is slowed down by its complex iterative TCC generation and decomposition.

\subsection{Ablation Study}
\begin{figure}[htp!]
  \centering
  \resizebox{\linewidth}{!}{
    \begin{tabular}{c}
      \subfloat[]{\input{figs/pgfplot/ICCAD_mean_std/ICCAD_mean_std}}
      \subfloat[]{\input{figs/pgfplot/ICCAD-L_mean_std/ICCAD-L_mean_std}}
    \end{tabular}
  }
  \caption{Mean and STD of (a) ICCAD (b) ICCAD-L datasets.}
  \label{fig:mean-std-of-bismo}
\end{figure}
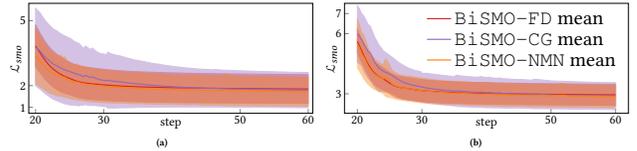

{\texttt{BiSMO-FD} vs. \texttt{BiSMO-NMN} vs. \texttt{BiSMO-CG}:}
\Cref{fig:loss_curves}, \Cref{tab:all_data}, and \Cref{tab:epe_tat} clearly show that \texttt{NMN} typically outperforms other methods, followed by \texttt{CG}, with \texttt{FD} being relatively weaker among all \texttt{BiSMO} variants.
The relative instability of \texttt{CG} is indicated by its largest standard deviation (STD) in \Cref{fig:mean-std-of-bismo}.
Meanwhile, \texttt{FD} boasts the shortest runtime (\Cref{tab:epe_tat}),
and \texttt{CG}'s advantage lies in outperforming \texttt{NMN} in some cases, as shown in \Cref{fig:ispd-test12-cg-better}.

%% file: doc/res/dataset.tex
\begin{table}[htp!]
	\centering
	\caption{Details of the Dataset.}
	\label{tab:data}
    \resizebox{.98\linewidth}{!}{
        \begin{tabular}{|l|l|ccccc|}
            \hline
            \multicolumn{1}{|c|}{Dataset}     & From &Area$^\dagger$ & Test num.  & Layer & CD$^\ddagger$           & tile \\ \hline \hline
            \textbf{ICCAD13}   & \cite{OPC-DAC2014-Gao,OPC-DAC2018-Yang}      & 202655  & 10  & Metal & 32$nm$  &4$\mu m^2$  \\
            \textbf{ICCAD-L} & \cite{OPC-ICCAD2020-NeuralILT,multiilt2023sun} & 475571   & 10  & Metal & 32$nm$  &4$\mu m^2$  \\
            \textbf{ISPD19}    & \cite{ispd2019-benchmark} & 698743  & 100   & Metal+Via & 28$nm$               &4$\mu m^2$\\
            \hline
            \multicolumn{6}{l}{\small{Area$^\dagger$: average area: unit $nm^2$; CD$^\ddagger$: critical dimension.}} \\
        \end{tabular}
        }
\end{table}

%% file: doc/res/all_data.tex
\begin{table*}[!htp]\centering
  \caption{Result comparison with SOTA.}\label{tab:all_data}
  \resizebox{.96\linewidth}{!}{
  \begin{tabular}{|l|cccccc|cccc|cccccc|}\hline
      \multirow{3}{*}{Bench} &\multicolumn{6}{c|}{MO} &\multicolumn{4}{c|}{\texttt{AM-SMO}} &\multicolumn{6}{c|}{\texttt{BiSMO} (Ours)} \\ \cline{2-17} 
  &\multicolumn{2}{c}{NILT~\cite{OPC-ICCAD2020-NeuralILT}} &\multicolumn{2}{c}{\small DAC23-MILT~\cite{multiilt2023sun}} &\multicolumn{2}{c|}{Abbe-MO(Ours)} &\multicolumn{2}{c}{\small Abbe-Hopkins$^\star$~\cite{ding2020gradient}} &\multicolumn{2}{c|}{\small Abbe-Abbe~\cite{sampleICC2022sun}} &\multicolumn{2}{c}{\texttt{BiSMO-FD}} &\multicolumn{2}{c}{\texttt{BiSMO-CG}} &\multicolumn{2}{c|}{\texttt{BiSMO-NMN}}\\
  &L2 &PVB &L2 &PVB &L2 &PVB &L2 &PVB &L2 &PVB &L2 &PVB &L2 &PVB &L2 &PVB \\\hline \hline 
  \textbf{ICCAD13} &37515 &50964 &28362 &40044 &20419 &29697 &27299 &37278 &17539 &23944 &13828 &17872 &13603 &16274  &\textbf{13059} &\textbf{15839} \\
  \textbf{ICCAD-L} &71570 &108162 &53143 &87010 &44478 &66092 &48879 &77062 &40455 &58560 &29779 &42643 &29762 &40543 &\textbf{28946} &\textbf{38706} \\
  \textbf{ISPD19} &97891 &119732 &85234 &105592 &61374 &93132 &79634 &97073 &55588 &84402 &39959 &64211 &39488 &61190 &\textbf{38737} &\textbf{59832} \\\hline
  Average &68992 &92953 &55580 &77549 &42090 &62974 &51937 &70471 &37861 &55635 &27855 &41576 &27618 &39336 &\textbf{26914} &\textbf{38126} \\
  Ratio &2.56 &2.44 &2.07 &2.03 &1.56 &1.65 &1.93 &1.85 &1.41 &1.46 &1.03 &1.09 &1.03 &1.03 &\textbf{1.00} &\textbf{1.00} \\
  \hline
  \multicolumn{16}{l}{\small{Abbe-Hopkins$^\star$~\cite{ding2020gradient}: \texttt{AM-SMO} employs Abbe model for SO and Hopkins model for MO. L2 and PVB unit: $nm^2$.}} \\
  \end{tabular}
  }
\end{table*}

%% file: doc/res/epe_runtime.tex
\begin{table}[!htp]\centering
  \caption{EPE and runtime comparison.}\label{tab:epe_tat}
  \resizebox{.96\linewidth}{!}{
  \begin{tabular}{|c|ccc|cc|ccc|}\hline
  \multirow{2}{*}{} &\multicolumn{3}{c|}{MO} &\multicolumn{2}{c|}{\texttt{AM-SMO}} & &\texttt{BiSMO} & \\\cline{2-9}
  &{\small \makecell{NILT\\\cite{OPC-ICCAD2020-NeuralILT}}} &{\makecell{\small DAC23\\\cite{multiilt2023sun}}} &\makecell{\small Abbe\\-MO} &\makecell{\small A$\sim$H$^\star$\\\cite{ding2020gradient}} &\makecell{\small A$\sim$A$\dagger$\\\cite{sampleICC2022sun}} &\texttt{FD} &\texttt{CG} & \texttt{NMN} \\\hline \hline
    EPE avg. &10.1 &3.6 &2.8 &3.3 &2.4 &1.8 &1.6 &\textbf{1.6} \\
    Ratio &6.2 &2.2 &1.7 &2.0 &1.5 &1.1 &1.0 &\textbf{1.0} \\\hline
    TAT$^\ddagger$ avg. &12.4 &3.8 &11.7 &287 &122.5 &\textbf{12.6} &15.3 &14.7 \\
    Ratio &0.84 &0.26 &0.80 &19.52 &8.33 &\textbf{0.86} &1.04 &1.00 \\
  \hline
  \multicolumn{9}{l}{\small{A$\sim$H$^\star$: Abbe-Hopkins;~A$\sim$A$^\dagger$: Abbe-Abbe;~TAT$^\ddagger$: Turn around time (s).}} \\
  \end{tabular}
  }
\end{table}

%% file: figs/pgfplot/ICCAD_mean_std/ICCAD_mean_std.tex
\pgfplotsset{
  width=\linewidth,
  height=.5\linewidth
}
\noindent
\begin{tikzpicture}[inner sep=2pt, outer sep=0pt]
  \definecolor{darkgray176}{RGB}{176,176,176}

  \definecolor{forestgreen4416044}{RGB}{44,160,44}
  \definecolor{lightgray204}{RGB}{204,204,204}
  \definecolor{steelblue31119180}{RGB}{31,119,180}
  \definecolor{crimson2143940}{RGB}{214,39,40}
  \definecolor{mediumpurple148103189}{RGB}{148,103,189}
  \definecolor{darkorange25512714}{RGB}{255,127,14}

  \begin{axis}[
      legend cell align={left},
      legend style={
          fill opacity=0.8,
          draw opacity=1,
          text opacity=1,
          draw=none,
          font=\huge
        },
      xmin=57, xmax=182,
      ymin=1.44414607004279, ymax=11.5992530193289,
      tick align=inside,
      tick pos=left,
      tick style={major tick length=3pt},
      xtick = {60, 90, 150, 180},
      xticklabels = {20, 30, 50, 60},
      ytick = {2, 4, 10},
      yticklabels = {1, 2, 5},
      xlabel = {step},
      xlabel near ticks,
      xlabel shift={-8pt},
      ylabel= {$\mathcal{L}_{smo}$},
      ylabel near ticks,
      ylabel shift={-5pt},
    ]
    \path [draw=crimson2143940, fill=crimson2143940, opacity=0.4]
    (axis cs:60,9.59268441634575)
    --(axis cs:60,5.73282284302314)
    --(axis cs:61,5.46622240747491)
    --(axis cs:62,5.19124546959169)
    --(axis cs:63,4.91595182917583)
    --(axis cs:64,4.65024888752238)
    --(axis cs:65,4.40392878024212)
    --(axis cs:66,4.18432616436645)
    --(axis cs:67,3.99437748812955)
    --(axis cs:68,3.83081499172733)
    --(axis cs:69,3.68627088981281)
    --(axis cs:70,3.55777588418497)
    --(axis cs:71,3.44808425292009)
    --(axis cs:72,3.35915630673613)
    --(axis cs:73,3.28748718877635)
    --(axis cs:74,3.22521297444799)
    --(axis cs:75,3.16309568847916)
    --(axis cs:76,3.09457367441412)
    --(axis cs:77,3.02043960977065)
    --(axis cs:78,2.94760480513279)
    --(axis cs:79,2.88320027009382)
    --(axis cs:80,2.83119223304466)
    --(axis cs:81,2.79206171011023)
    --(axis cs:82,2.76381968559328)
    --(axis cs:83,2.74329276770571)
    --(axis cs:84,2.72721981191369)
    --(axis cs:85,2.7129825403483)
    --(axis cs:86,2.69893057310163)
    --(axis cs:87,2.68436007120336)
    --(axis cs:88,2.66924470532819)
    --(axis cs:89,2.65391031198687)
    --(axis cs:90,2.63878604859468)
    --(axis cs:91,2.62428314104667)
    --(axis cs:92,2.6107406309502)
    --(axis cs:93,2.59838480668457)
    --(axis cs:94,2.58729862425099)
    --(axis cs:95,2.57741430116298)
    --(axis cs:96,2.56854798519805)
    --(axis cs:97,2.56046011087278)
    --(axis cs:98,2.55291487961723)
    --(axis cs:99,2.54571610000992)
    --(axis cs:100,2.53872960860905)
    --(axis cs:101,2.53189111549484)
    --(axis cs:102,2.52520181977437)
    --(axis cs:103,2.51870702021839)
    --(axis cs:104,2.51247247090212)
    --(axis cs:105,2.50655676185099)
    --(axis cs:106,2.50099697525849)
    --(axis cs:107,2.4958015638026)
    --(axis cs:108,2.490953091028)
    --(axis cs:109,2.48641250924587)
    --(axis cs:110,2.48212738467051)
    --(axis cs:111,2.47803984566259)
    --(axis cs:112,2.47409524858308)
    --(axis cs:113,2.4702511954345)
    --(axis cs:114,2.46648453981527)
    --(axis cs:115,2.46279357684993)
    --(axis cs:116,2.45919313422173)
    --(axis cs:117,2.45570552962472)
    --(axis cs:118,2.45235274233007)
    --(axis cs:119,2.44914940635306)
    --(axis cs:120,2.44609947846648)
    --(axis cs:121,2.44319678709029)
    --(axis cs:122,2.44042625544832)
    --(axis cs:123,2.4377692426676)
    --(axis cs:124,2.43520700573806)
    --(axis cs:125,2.43272636183809)
    --(axis cs:126,2.43032204181041)
    --(axis cs:127,2.42799535720126)
    --(axis cs:128,2.42575099656704)
    --(axis cs:129,2.42359433491566)
    --(axis cs:130,2.42152965935968)
    --(axis cs:131,2.41955811610011)
    --(axis cs:132,2.41767571003434)
    --(axis cs:133,2.41587474908486)
    --(axis cs:134,2.41414401846268)
    --(axis cs:135,2.4124714111715)
    --(axis cs:136,2.41084470191729)
    --(axis cs:137,2.40925539852852)
    --(axis cs:138,2.40770011343795)
    --(axis cs:139,2.4061824980334)
    --(axis cs:140,2.40471161518176)
    --(axis cs:141,2.40329633228676)
    --(axis cs:142,2.40193919952211)
    --(axis cs:143,2.40063309140621)
    --(axis cs:144,2.39936427351562)
    --(axis cs:145,2.39811379887128)
    --(axis cs:146,2.39686434968852)
    --(axis cs:147,2.39560519832803)
    --(axis cs:148,2.39433630304658)
    --(axis cs:149,2.39306594333739)
    --(axis cs:150,2.39180410447735)
    --(axis cs:151,2.39055914106133)
    --(axis cs:152,2.38933826204416)
    --(axis cs:153,2.38814543260066)
    --(axis cs:154,2.38698274016258)
    --(axis cs:155,2.38585110159992)
    --(axis cs:156,2.38475170268399)
    --(axis cs:157,2.38368518103016)
    --(axis cs:158,2.38265173723634)
    --(axis cs:159,2.38165063837519)
    --(axis cs:160,2.3806812356144)
    --(axis cs:161,2.37974296336693)
    --(axis cs:162,2.37883557074777)
    --(axis cs:163,2.3779588135642)
    --(axis cs:164,2.37711218474856)
    --(axis cs:165,2.37629429618815)
    --(axis cs:166,2.3755022225896)
    --(axis cs:167,2.37473361565038)
    --(axis cs:168,2.37398424834048)
    --(axis cs:169,2.37325107683634)
    --(axis cs:170,2.37253144556347)
    --(axis cs:171,2.37182373166227)
    --(axis cs:172,2.37112777757729)
    --(axis cs:173,2.37044219770736)
    --(axis cs:174,2.3697679809625)
    --(axis cs:175,2.36910539551531)
    --(axis cs:176,2.36845508407149)
    --(axis cs:177,2.36781732414397)
    --(axis cs:178,2.36719150558841)
    --(axis cs:179,2.36657708927904)
    --(axis cs:180,2.36597271440885)
    --(axis cs:180,5.01830071927645)
    --(axis cs:180,5.01830071927645)
    --(axis cs:179,5.01966401293959)
    --(axis cs:178,5.02103219016659)
    --(axis cs:177,5.02240975524751)
    --(axis cs:176,5.02380153347459)
    --(axis cs:175,5.02521349505629)
    --(axis cs:174,5.02664634665896)
    --(axis cs:173,5.02810342511826)
    --(axis cs:172,5.02958423566734)
    --(axis cs:171,5.03108893275118)
    --(axis cs:170,5.03262099450764)
    --(axis cs:169,5.03419008622671)
    --(axis cs:168,5.03580269128049)
    --(axis cs:167,5.03746759818629)
    --(axis cs:166,5.03918820042218)
    --(axis cs:165,5.04096656520864)
    --(axis cs:164,5.04280642391691)
    --(axis cs:163,5.04470487509547)
    --(axis cs:162,5.04665859944113)
    --(axis cs:161,5.04866194004494)
    --(axis cs:160,5.05070819157066)
    --(axis cs:159,5.05279002210149)
    --(axis cs:158,5.0548980922467)
    --(axis cs:157,5.05702629338848)
    --(axis cs:156,5.05917010174411)
    --(axis cs:155,5.06132945565106)
    --(axis cs:154,5.06350844979408)
    --(axis cs:153,5.06571463572057)
    --(axis cs:152,5.06795219663503)
    --(axis cs:151,5.07022574720618)
    --(axis cs:150,5.07254337725025)
    --(axis cs:149,5.07491544284731)
    --(axis cs:148,5.07735416097319)
    --(axis cs:147,5.07987535816001)
    --(axis cs:146,5.0824859310732)
    --(axis cs:145,5.08518984340167)
    --(axis cs:144,5.08797946694764)
    --(axis cs:143,5.09083366922916)
    --(axis cs:142,5.09372439567747)
    --(axis cs:141,5.09662627704246)
    --(axis cs:140,5.09953817214887)
    --(axis cs:139,5.10249658197606)
    --(axis cs:138,5.10557169518632)
    --(axis cs:137,5.10882910369163)
    --(axis cs:136,5.11230133136976)
    --(axis cs:135,5.11597795730537)
    --(axis cs:134,5.11983034295699)
    --(axis cs:133,5.12382354174003)
    --(axis cs:132,5.12792568175796)
    --(axis cs:131,5.13211164841863)
    --(axis cs:130,5.13636795607305)
    --(axis cs:129,5.14069327990672)
    --(axis cs:128,5.14509601404545)
    --(axis cs:127,5.1495832112573)
    --(axis cs:126,5.15414916008626)
    --(axis cs:125,5.15878803482939)
    --(axis cs:124,5.16351042151566)
    --(axis cs:123,5.16835184440668)
    --(axis cs:122,5.17335083367064)
    --(axis cs:121,5.17852591112138)
    --(axis cs:120,5.18387386156801)
    --(axis cs:119,5.18938718994516)
    --(axis cs:118,5.1950591047845)
    --(axis cs:117,5.2008848367388)
    --(axis cs:116,5.20685706147224)
    --(axis cs:115,5.21297563301182)
    --(axis cs:114,5.21925330846659)
    --(axis cs:113,5.22572015046699)
    --(axis cs:112,5.23242435514617)
    --(axis cs:111,5.23942455367518)
    --(axis cs:110,5.24677831696676)
    --(axis cs:109,5.25452815695286)
    --(axis cs:108,5.2626934972975)
    --(axis cs:107,5.27126644080555)
    --(axis cs:106,5.28021735732207)
    --(axis cs:105,5.28950432480367)
    --(axis cs:104,5.29910275111326)
    --(axis cs:103,5.30903150326488)
    --(axis cs:102,5.31937945044353)
    --(axis cs:101,5.3303143660344)
    --(axis cs:100,5.34206107799831)
    --(axis cs:99,5.35484346839523)
    --(axis cs:98,5.36882877558755)
    --(axis cs:97,5.38407061460635)
    --(axis cs:96,5.40048563754365)
    --(axis cs:95,5.41786105322239)
    --(axis cs:94,5.43591297414531)
    --(axis cs:93,5.45435838980285)
    --(axis cs:92,5.47303572383954)
    --(axis cs:91,5.49201562985787)
    --(axis cs:90,5.51162910967711)
    --(axis cs:89,5.53236140317732)
    --(axis cs:88,5.55463985334949)
    --(axis cs:87,5.57862353704249)
    --(axis cs:86,5.60410804784716)
    --(axis cs:85,5.63064064480948)
    --(axis cs:84,5.65781467771797)
    --(axis cs:83,5.68567151337645)
    --(axis cs:82,5.71508480010923)
    --(axis cs:81,5.74801588083215)
    --(axis cs:80,5.78747131637856)
    --(axis cs:79,5.8369853912221)
    --(axis cs:78,5.89960402856166)
    --(axis cs:77,5.97664119314683)
    --(axis cs:76,6.06686703183894)
    --(axis cs:75,6.16704137359359)
    --(axis cs:74,6.2743867207577)
    --(axis cs:73,6.38966940263905)
    --(axis cs:72,6.51594592715059)
    --(axis cs:71,6.65394325867659)
    --(axis cs:70,6.80268201300131)
    --(axis cs:69,6.96419879478802)
    --(axis cs:68,7.14358581469967)
    --(axis cs:67,7.34627454853732)
    --(axis cs:66,7.57810536181763)
    --(axis cs:65,7.8443897488869)
    --(axis cs:64,8.14776532413229)
    --(axis cs:63,8.48462881543172)
    --(axis cs:62,8.8457740692877)
    --(axis cs:61,9.2191098908039)
    --(axis cs:60,9.59268441634575)
    --cycle;

    \path [draw=mediumpurple148103189, fill=mediumpurple148103189, opacity=0.4]
    (axis cs:60,11.1376572489068)
    --(axis cs:60,3.97798024422672)
    --(axis cs:61,3.81683053306536)
    --(axis cs:62,3.67625163992746)
    --(axis cs:63,3.48884467088813)
    --(axis cs:64,3.29940508162977)
    --(axis cs:65,3.14357536579188)
    --(axis cs:66,2.99106152082212)
    --(axis cs:67,2.84861284986926)
    --(axis cs:68,2.71791703429961)
    --(axis cs:69,2.64610183823525)
    --(axis cs:70,2.5498048239241)
    --(axis cs:71,2.46053080350704)
    --(axis cs:72,2.38931250819685)
    --(axis cs:73,2.31329913799222)
    --(axis cs:74,2.26018124318193)
    --(axis cs:75,2.21860411467917)
    --(axis cs:76,2.19813493622656)
    --(axis cs:77,2.19288524276483)
    --(axis cs:78,2.1642527027915)
    --(axis cs:79,2.12314126966425)
    --(axis cs:80,2.19781758030651)
    --(axis cs:81,2.21021095980583)
    --(axis cs:82,2.23886286258456)
    --(axis cs:83,2.19861985016293)
    --(axis cs:84,2.16535992776659)
    --(axis cs:85,2.16103415647631)
    --(axis cs:86,2.16424501577714)
    --(axis cs:87,2.13914425778865)
    --(axis cs:88,2.12224179942593)
    --(axis cs:89,2.10230393273131)
    --(axis cs:90,2.05578449652904)
    --(axis cs:91,2.05606271447886)
    --(axis cs:92,1.90574184046488)
    --(axis cs:93,1.9390686010851)
    --(axis cs:94,1.93297301886717)
    --(axis cs:95,1.93472041696717)
    --(axis cs:96,1.93600593358005)
    --(axis cs:97,1.9551232415222)
    --(axis cs:98,1.95642297574667)
    --(axis cs:99,1.95266642925744)
    --(axis cs:100,1.95535367268414)
    --(axis cs:101,1.96108006977115)
    --(axis cs:102,1.96514505228879)
    --(axis cs:103,1.96934551442649)
    --(axis cs:104,1.97349645216253)
    --(axis cs:105,1.99285431829261)
    --(axis cs:106,1.99259437323806)
    --(axis cs:107,1.99722019141956)
    --(axis cs:108,2.00237343713435)
    --(axis cs:109,2.00312915516131)
    --(axis cs:110,2.00461985721196)
    --(axis cs:111,2.00658771728992)
    --(axis cs:112,2.00871643851952)
    --(axis cs:113,2.01040273607064)
    --(axis cs:114,2.00125015829361)
    --(axis cs:115,2.00885828771668)
    --(axis cs:116,2.01127434984982)
    --(axis cs:117,2.01396818885557)
    --(axis cs:118,2.01600631620085)
    --(axis cs:119,2.01770477662715)
    --(axis cs:120,2.01898898306897)
    --(axis cs:121,2.01995810904689)
    --(axis cs:122,2.02068327194843)
    --(axis cs:123,2.02135304178759)
    --(axis cs:124,2.02196403198455)
    --(axis cs:125,2.02252190730897)
    --(axis cs:126,2.02244747732175)
    --(axis cs:127,2.02282429392511)
    --(axis cs:128,2.02290883897114)
    --(axis cs:129,2.02240673258702)
    --(axis cs:130,2.02227807969913)
    --(axis cs:131,2.02199809511794)
    --(axis cs:132,2.02158778517985)
    --(axis cs:133,2.02104066675506)
    --(axis cs:134,2.02034974718471)
    --(axis cs:135,2.01950531188128)
    --(axis cs:136,2.01853863543449)
    --(axis cs:137,2.01811051584982)
    --(axis cs:138,2.01735730392906)
    --(axis cs:139,2.03177081841269)
    --(axis cs:140,2.02264087253694)
    --(axis cs:141,2.01929507449862)
    --(axis cs:142,2.01673826745562)
    --(axis cs:143,2.01362374858443)
    --(axis cs:144,2.01115831284598)
    --(axis cs:145,2.00872267766521)
    --(axis cs:146,2.0068157014939)
    --(axis cs:147,2.00481554508509)
    --(axis cs:148,2.00248772295986)
    --(axis cs:149,2.00094890823481)
    --(axis cs:150,1.99901135954216)
    --(axis cs:151,1.99715836793751)
    --(axis cs:152,1.99529768839057)
    --(axis cs:153,1.9934726244857)
    --(axis cs:154,1.99169057397749)
    --(axis cs:155,1.98996989513403)
    --(axis cs:156,2.02885279918474)
    --(axis cs:157,2.02071338531282)
    --(axis cs:158,2.01326341878074)
    --(axis cs:159,2.00538471549938)
    --(axis cs:160,1.99828704230483)
    --(axis cs:161,1.99251937822482)
    --(axis cs:162,1.9880852641667)
    --(axis cs:163,1.98485751192568)
    --(axis cs:164,1.9825631611638)
    --(axis cs:165,1.98078470044776)
    --(axis cs:166,1.97890081659788)
    --(axis cs:167,1.97727389918852)
    --(axis cs:168,1.97550203308947)
    --(axis cs:169,1.97437833277958)
    --(axis cs:170,1.9718788224006)
    --(axis cs:171,1.96930967732012)
    --(axis cs:172,1.96703391028345)
    --(axis cs:173,1.96504046106606)
    --(axis cs:174,1.9631902826032)
    --(axis cs:175,1.96151297340358)
    --(axis cs:176,1.96021775449259)
    --(axis cs:177,1.95869670269773)
    --(axis cs:178,1.9572173806565)
    --(axis cs:179,1.95574870161165)
    --(axis cs:180,1.96075776744113)
    --(axis cs:180,5.22158240151181)
    --(axis cs:180,5.22158240151181)
    --(axis cs:179,5.21079127736937)
    --(axis cs:178,5.21594043050836)
    --(axis cs:177,5.22111045958712)
    --(axis cs:176,5.2261739329006)
    --(axis cs:175,5.22787856807743)
    --(axis cs:174,5.23200598833949)
    --(axis cs:173,5.23625763272972)
    --(axis cs:172,5.24054598855077)
    --(axis cs:171,5.24676056937635)
    --(axis cs:170,5.2559865159249)
    --(axis cs:169,5.26568292649014)
    --(axis cs:168,5.26204996600263)
    --(axis cs:167,5.26733149898959)
    --(axis cs:166,5.27270436032778)
    --(axis cs:165,5.28103594965295)
    --(axis cs:164,5.28638975688888)
    --(axis cs:163,5.29176435430052)
    --(axis cs:162,5.29726133922686)
    --(axis cs:161,5.30276658578733)
    --(axis cs:160,5.30809806950395)
    --(axis cs:159,5.3133922019582)
    --(axis cs:158,5.31879219759805)
    --(axis cs:157,5.32445716026518)
    --(axis cs:156,5.33033072685438)
    --(axis cs:155,5.3406892797851)
    --(axis cs:154,5.34748202772234)
    --(axis cs:153,5.35443885260323)
    --(axis cs:152,5.36149047002618)
    --(axis cs:151,5.36859733312756)
    --(axis cs:150,5.3756281754081)
    --(axis cs:149,5.38287069568517)
    --(axis cs:148,5.38894854393925)
    --(axis cs:147,5.39737075328527)
    --(axis cs:146,5.40491718672784)
    --(axis cs:145,5.41238013224079)
    --(axis cs:144,5.42166746706887)
    --(axis cs:143,5.43140582008775)
    --(axis cs:142,5.44359615274854)
    --(axis cs:141,5.45468697353605)
    --(axis cs:140,5.4692409510028)
    --(axis cs:139,5.50807510119638)
    --(axis cs:138,5.46492814795997)
    --(axis cs:137,5.47453627370096)
    --(axis cs:136,5.48438969307961)
    --(axis cs:135,5.49665303048017)
    --(axis cs:134,5.50967673635105)
    --(axis cs:133,5.52342418367066)
    --(axis cs:132,5.53795642048574)
    --(axis cs:131,5.55334029237138)
    --(axis cs:130,5.56962842016354)
    --(axis cs:129,5.58687022969326)
    --(axis cs:128,5.60487949015331)
    --(axis cs:127,5.62498061005896)
    --(axis cs:126,5.64464441682803)
    --(axis cs:125,5.67342421390686)
    --(axis cs:124,5.69531969375398)
    --(axis cs:123,5.71830910478071)
    --(axis cs:122,5.74228400462475)
    --(axis cs:121,5.76792960724644)
    --(axis cs:120,5.79419325169513)
    --(axis cs:119,5.82179927458134)
    --(axis cs:118,5.85138727281892)
    --(axis cs:117,5.88238069286592)
    --(axis cs:116,5.91575905068576)
    --(axis cs:115,5.94900611600799)
    --(axis cs:114,5.99622024442398)
    --(axis cs:113,6.0022200494976)
    --(axis cs:112,6.03860266376777)
    --(axis cs:111,6.07818508410931)
    --(axis cs:110,6.11854670391475)
    --(axis cs:109,6.16061212348707)
    --(axis cs:108,6.20574006632176)
    --(axis cs:107,6.24296776348308)
    --(axis cs:106,6.28428333996537)
    --(axis cs:105,6.33955484899712)
    --(axis cs:104,6.35758632104609)
    --(axis cs:103,6.40928123747323)
    --(axis cs:102,6.46370655217288)
    --(axis cs:101,6.5204268977734)
    --(axis cs:100,6.57355121832996)
    --(axis cs:99,6.63299140574927)
    --(axis cs:98,6.69845239809367)
    --(axis cs:97,6.76552011194002)
    --(axis cs:96,6.78755580634151)
    --(axis cs:95,6.84287872701477)
    --(axis cs:94,6.90310459973177)
    --(axis cs:93,6.95242495779857)
    --(axis cs:92,7.10094940079946)
    --(axis cs:91,6.83652745065939)
    --(axis cs:90,6.93042229725605)
    --(axis cs:89,6.93156802790865)
    --(axis cs:88,7.0131883696796)
    --(axis cs:87,7.12937945913793)
    --(axis cs:86,7.27557868322036)
    --(axis cs:85,7.47528488477583)
    --(axis cs:84,7.32959515894147)
    --(axis cs:83,7.49962991904788)
    --(axis cs:82,7.65355714321378)
    --(axis cs:81,7.41622606049599)
    --(axis cs:80,7.70877488414051)
    --(axis cs:79,7.9014058780866)
    --(axis cs:78,7.98602300445337)
    --(axis cs:77,8.09490033500922)
    --(axis cs:76,8.25420857535486)
    --(axis cs:75,8.38484022316568)
    --(axis cs:74,8.47576264643599)
    --(axis cs:73,8.5740960532004)
    --(axis cs:72,8.68698491802691)
    --(axis cs:71,8.81014332979374)
    --(axis cs:70,8.95774279303452)
    --(axis cs:69,9.11155673873008)
    --(axis cs:68,9.19409907135224)
    --(axis cs:67,9.3861942599635)
    --(axis cs:66,9.60185291742556)
    --(axis cs:65,9.83102108692113)
    --(axis cs:64,10.0717855616808)
    --(axis cs:63,10.3882386306561)
    --(axis cs:62,10.7124799350847)
    --(axis cs:61,10.9185503072362)
    --(axis cs:60,11.1376572489068)
    --cycle;

    \path [draw=darkorange25512714, fill=darkorange25512714, opacity=0.4]
    (axis cs:60,9.24199351500348)
    --(axis cs:60,4.18448720742389)
    --(axis cs:61,4.02409284392272)
    --(axis cs:62,3.8614448041941)
    --(axis cs:63,3.76467432857022)
    --(axis cs:64,3.61262187257799)
    --(axis cs:65,3.46267635250778)
    --(axis cs:66,3.33815788375726)
    --(axis cs:67,3.25435027385398)
    --(axis cs:68,3.18123276267392)
    --(axis cs:69,3.09511184498171)
    --(axis cs:70,3.02749954407076)
    --(axis cs:71,3.01786706613868)
    --(axis cs:72,2.97661657463959)
    --(axis cs:73,2.91438942147011)
    --(axis cs:74,2.8719136212433)
    --(axis cs:75,2.81572803780378)
    --(axis cs:76,2.76190264904206)
    --(axis cs:77,2.71365050870286)
    --(axis cs:78,2.69295890737566)
    --(axis cs:79,2.66121981743255)
    --(axis cs:80,2.63319012963141)
    --(axis cs:81,2.60252100965385)
    --(axis cs:82,2.57678373002184)
    --(axis cs:83,2.55382558211706)
    --(axis cs:84,2.53424251562732)
    --(axis cs:85,2.51518588711733)
    --(axis cs:86,2.49813546077626)
    --(axis cs:87,2.48203442676334)
    --(axis cs:88,2.4671767158877)
    --(axis cs:89,2.45362081190393)
    --(axis cs:90,2.44112867567808)
    --(axis cs:91,2.42931815682941)
    --(axis cs:92,2.41771549558877)
    --(axis cs:93,2.40774335881798)
    --(axis cs:94,2.39765433365595)
    --(axis cs:95,2.38799960277249)
    --(axis cs:96,2.37903859803531)
    --(axis cs:97,2.39879523907602)
    --(axis cs:98,2.38874973891895)
    --(axis cs:99,2.37742403250796)
    --(axis cs:100,2.3663485531544)
    --(axis cs:101,2.35597145323997)
    --(axis cs:102,2.34657103997967)
    --(axis cs:103,2.33850702500635)
    --(axis cs:104,2.33417317750632)
    --(axis cs:105,2.32786399515503)
    --(axis cs:106,2.34188491426284)
    --(axis cs:107,2.32191228289037)
    --(axis cs:108,2.318530202537)
    --(axis cs:109,2.31035718982883)
    --(axis cs:110,2.30495052653497)
    --(axis cs:111,2.29941186848211)
    --(axis cs:112,2.29429338784413)
    --(axis cs:113,2.28936487923541)
    --(axis cs:114,2.28442291150592)
    --(axis cs:115,2.27987145203248)
    --(axis cs:116,2.27455657489205)
    --(axis cs:117,2.26860482768917)
    --(axis cs:118,2.26289162528686)
    --(axis cs:119,2.25720171175155)
    --(axis cs:120,2.25191002432329)
    --(axis cs:121,2.26944495212268)
    --(axis cs:122,2.2599752289289)
    --(axis cs:123,2.25241191949921)
    --(axis cs:124,2.2442627726094)
    --(axis cs:125,2.2368505396354)
    --(axis cs:126,2.23051920603691)
    --(axis cs:127,2.22523778675177)
    --(axis cs:128,2.23834170077499)
    --(axis cs:129,2.25570596082187)
    --(axis cs:130,2.24684895337034)
    --(axis cs:131,2.23619049549907)
    --(axis cs:132,2.22499937357966)
    --(axis cs:133,2.21462036508692)
    --(axis cs:134,2.20604047540768)
    --(axis cs:135,2.19966427963475)
    --(axis cs:136,2.19529355459688)
    --(axis cs:137,2.19236476301454)
    --(axis cs:138,2.19874897221909)
    --(axis cs:139,2.19492388813969)
    --(axis cs:140,2.19299240916357)
    --(axis cs:141,2.18865925692894)
    --(axis cs:142,2.18451097370627)
    --(axis cs:143,2.15340767010506)
    --(axis cs:144,2.1524600384437)
    --(axis cs:145,2.14651073952065)
    --(axis cs:146,2.16290990357881)
    --(axis cs:147,2.16101578783582)
    --(axis cs:148,2.15942404783323)
    --(axis cs:149,2.15725962998862)
    --(axis cs:150,2.15508246745016)
    --(axis cs:151,2.15316761924244)
    --(axis cs:152,2.1512636956532)
    --(axis cs:153,2.1493400464033)
    --(axis cs:154,2.14735156433678)
    --(axis cs:155,2.14526848974177)
    --(axis cs:156,2.14309155547558)
    --(axis cs:157,2.14084754451058)
    --(axis cs:158,2.13857972041468)
    --(axis cs:159,2.13633976807225)
    --(axis cs:160,2.13417666343931)
    --(axis cs:161,2.14103177691077)
    --(axis cs:162,2.13806184573883)
    --(axis cs:163,2.13442954167838)
    --(axis cs:164,2.13083227126559)
    --(axis cs:165,2.12747816801071)
    --(axis cs:166,2.12455510925384)
    --(axis cs:167,2.12213348455151)
    --(axis cs:168,2.12018514692815)
    --(axis cs:169,2.11854013781995)
    --(axis cs:170,2.11709935413015)
    --(axis cs:171,2.11571053206061)
    --(axis cs:172,2.11427214080133)
    --(axis cs:173,2.11272322737477)
    --(axis cs:174,2.11104536104846)
    --(axis cs:175,2.10925264756857)
    --(axis cs:176,2.1073795305573)
    --(axis cs:177,2.10597650570879)
    --(axis cs:178,2.10373672605409)
    --(axis cs:179,2.10176045615244)
    --(axis cs:180,2.09976543920568)
    --(axis cs:180,4.75132985574671)
    --(axis cs:180,4.75132985574671)
    --(axis cs:179,4.75502620022726)
    --(axis cs:178,4.75795417665587)
    --(axis cs:177,4.76420252757062)
    --(axis cs:176,4.76760254036555)
    --(axis cs:175,4.770989652465)
    --(axis cs:174,4.77436053704571)
    --(axis cs:173,4.77769320405223)
    --(axis cs:172,4.78097395962439)
    --(axis cs:171,4.78423212827112)
    --(axis cs:170,4.78754096283305)
    --(axis cs:169,4.79111495632678)
    --(axis cs:168,4.79432546079127)
    --(axis cs:167,4.79941576414386)
    --(axis cs:166,4.80408551860719)
    --(axis cs:165,4.80931904554368)
    --(axis cs:164,4.81505386383573)
    --(axis cs:163,4.82108047857766)
    --(axis cs:162,4.82706505016571)
    --(axis cs:161,4.83245073651697)
    --(axis cs:160,4.82412724109408)
    --(axis cs:159,4.82765609393489)
    --(axis cs:158,4.83103206738135)
    --(axis cs:157,4.83422495857932)
    --(axis cs:156,4.83723326122821)
    --(axis cs:155,4.84010519800237)
    --(axis cs:154,4.84293783767128)
    --(axis cs:153,4.84590669775259)
    --(axis cs:152,4.84931380182909)
    --(axis cs:151,4.85347851322674)
    --(axis cs:150,4.85878078614328)
    --(axis cs:149,4.86644542334085)
    --(axis cs:148,4.8760444112961)
    --(axis cs:147,4.89131098676135)
    --(axis cs:146,4.9069738482904)
    --(axis cs:145,5.01686399440422)
    --(axis cs:144,5.01624418518911)
    --(axis cs:143,5.03259057894889)
    --(axis cs:142,4.89393572448252)
    --(axis cs:141,4.90420649624489)
    --(axis cs:140,4.91629535347834)
    --(axis cs:139,4.91854487807277)
    --(axis cs:138,4.93091377993239)
    --(axis cs:137,4.89765899301269)
    --(axis cs:136,4.90202395028593)
    --(axis cs:135,4.90853290874263)
    --(axis cs:134,4.91819582219974)
    --(axis cs:133,4.93184660535681)
    --(axis cs:132,4.94972985444005)
    --(axis cs:131,4.97152766226918)
    --(axis cs:130,4.9951168173511)
    --(axis cs:129,5.01643308775448)
    --(axis cs:128,4.97199231411759)
    --(axis cs:127,4.95005346538446)
    --(axis cs:126,4.95602542687478)
    --(axis cs:125,4.96201518397452)
    --(axis cs:124,4.96831752676804)
    --(axis cs:123,4.97549375924988)
    --(axis cs:122,4.98279372630182)
    --(axis cs:121,4.99159869182874)
    --(axis cs:120,5.00371591504591)
    --(axis cs:119,5.01508048333969)
    --(axis cs:118,5.03373568165132)
    --(axis cs:117,5.04580140037632)
    --(axis cs:116,5.06419068806266)
    --(axis cs:115,5.07423571330413)
    --(axis cs:114,5.07456759085156)
    --(axis cs:113,5.084001071349)
    --(axis cs:112,5.0932080665092)
    --(axis cs:111,5.10303442534876)
    --(axis cs:110,5.11429233235175)
    --(axis cs:109,5.12587592536739)
    --(axis cs:108,5.14225597319084)
    --(axis cs:107,5.15121746640773)
    --(axis cs:106,5.18851967729753)
    --(axis cs:105,5.16979709474213)
    --(axis cs:104,5.18315425035776)
    --(axis cs:103,5.18957572722143)
    --(axis cs:102,5.20552811640002)
    --(axis cs:101,5.22479446167702)
    --(axis cs:100,5.24695692017324)
    --(axis cs:99,5.27219826477903)
    --(axis cs:98,5.30295665623028)
    --(axis cs:97,5.32925732935965)
    --(axis cs:96,5.31186580469754)
    --(axis cs:95,5.32952337601016)
    --(axis cs:94,5.35182259982336)
    --(axis cs:93,5.37563824633033)
    --(axis cs:92,5.40090351724387)
    --(axis cs:91,5.42727786005444)
    --(axis cs:90,5.45462017800539)
    --(axis cs:89,5.4827126775141)
    --(axis cs:88,5.51159575744756)
    --(axis cs:87,5.54032399074765)
    --(axis cs:86,5.56967469318492)
    --(axis cs:85,5.60283622096068)
    --(axis cs:84,5.63100559228284)
    --(axis cs:83,5.67492068901636)
    --(axis cs:82,5.71561405516493)
    --(axis cs:81,5.75608411291237)
    --(axis cs:80,5.82427220500146)
    --(axis cs:79,5.91230694648347)
    --(axis cs:78,6.01422388147321)
    --(axis cs:77,5.91207118433608)
    --(axis cs:76,6.00734278476531)
    --(axis cs:75,6.12434344962297)
    --(axis cs:74,6.22641556519622)
    --(axis cs:73,6.64168288516288)
    --(axis cs:72,6.60104961264679)
    --(axis cs:71,6.84719488454491)
    --(axis cs:70,6.71608303840299)
    --(axis cs:69,6.83208022311826)
    --(axis cs:68,7.00311186279911)
    --(axis cs:67,7.13587713932351)
    --(axis cs:66,7.31407757652412)
    --(axis cs:65,7.57059262370377)
    --(axis cs:64,7.90942369207349)
    --(axis cs:63,8.25913691639438)
    --(axis cs:62,8.51581402110802)
    --(axis cs:61,8.8763831349286)
    --(axis cs:60,9.24199351500348)
    --cycle;

    \addplot [thick, crimson2143940]
    table {%
        60 7.66275362968445
        61 7.3426661491394
        62 7.0185097694397
        63 6.70029032230377
        64 6.39900710582733
        65 6.12415926456451
        66 5.88121576309204
        67 5.67032601833344
        68 5.4872004032135
        69 5.32523484230042
        70 5.18022894859314
        71 5.05101375579834
        72 4.93755111694336
        73 4.8385782957077
        74 4.74979984760284
        75 4.66506853103638
        76 4.58072035312653
        77 4.49854040145874
        78 4.42360441684723
        79 4.36009283065796
        80 4.30933177471161
        81 4.27003879547119
        82 4.23945224285126
        83 4.21448214054108
        84 4.19251724481583
        85 4.17181159257889
        86 4.1515193104744
        87 4.13149180412293
        88 4.11194227933884
        89 4.09313585758209
        90 4.07520757913589
        91 4.05814938545227
        92 4.04188817739487
        93 4.02637159824371
        94 4.01160579919815
        95 3.99763767719269
        96 3.98451681137085
        97 3.97226536273956
        98 3.96087182760239
        99 3.95027978420258
        100 3.94039534330368
        101 3.93110274076462
        102 3.92229063510895
        103 3.91386926174164
        104 3.90578761100769
        105 3.89803054332733
        106 3.89060716629028
        107 3.88353400230408
        108 3.87682329416275
        109 3.87047033309937
        110 3.86445285081863
        111 3.85873219966888
        112 3.85325980186462
        113 3.84798567295074
        114 3.84286892414093
        115 3.83788460493088
        116 3.83302509784698
        117 3.82829518318176
        118 3.82370592355728
        119 3.81926829814911
        120 3.81498667001724
        121 3.81086134910583
        122 3.80688854455948
        123 3.80306054353714
        124 3.79935871362686
        125 3.79575719833374
        126 3.79223560094833
        127 3.78878928422928
        128 3.78542350530624
        129 3.78214380741119
        130 3.77894880771637
        131 3.77583488225937
        132 3.77280069589615
        133 3.76984914541244
        134 3.76698718070984
        135 3.76422468423843
        136 3.76157301664352
        137 3.75904225111008
        138 3.75663590431213
        139 3.75433954000473
        140 3.75212489366531
        141 3.74996130466461
        142 3.74783179759979
        143 3.74573338031769
        144 3.74367187023163
        145 3.74165182113647
        146 3.73967514038086
        147 3.73774027824402
        148 3.73584523200989
        149 3.73399069309235
        150 3.7321737408638
        151 3.73039244413376
        152 3.7286452293396
        153 3.72693003416061
        154 3.72524559497833
        155 3.72359027862549
        156 3.72196090221405
        157 3.72035573720932
        158 3.71877491474152
        159 3.71722033023834
        160 3.71569471359253
        161 3.71420245170593
        162 3.71274708509445
        163 3.71133184432983
        164 3.70995930433273
        165 3.70863043069839
        166 3.70734521150589
        167 3.70610060691833
        168 3.70489346981049
        169 3.70372058153152
        170 3.70257622003555
        171 3.70145633220673
        172 3.70035600662231
        173 3.69927281141281
        174 3.69820716381073
        175 3.6971594452858
        176 3.69612830877304
        177 3.69511353969574
        178 3.6941118478775
        179 3.69312055110931
        180 3.69213671684265
      };
    \addplot [thick, mediumpurple148103189]
    table {%
        60 7.55781874656677
        61 7.36769042015076
        62 7.1943657875061
        63 6.93854165077209
        64 6.68559532165527
        65 6.48729822635651
        66 6.29645721912384
        67 6.11740355491638
        68 5.95600805282593
        69 5.87882928848267
        70 5.75377380847931
        71 5.63533706665039
        72 5.53814871311188
        73 5.44369759559631
        74 5.36797194480896
        75 5.30172216892242
        76 5.22617175579071
        77 5.14389278888702
        78 5.07513785362244
        79 5.01227357387543
        80 4.95329623222351
        81 4.81321851015091
        82 4.94621000289917
        83 4.84912488460541
        84 4.74747754335403
        85 4.81815952062607
        86 4.71991184949875
        87 4.63426185846329
        88 4.56771508455276
        89 4.51693598031998
        90 4.49310339689255
        91 4.44629508256912
        92 4.50334562063217
        93 4.44574677944183
        94 4.41803880929947
        95 4.38879957199097
        96 4.36178086996078
        97 4.36032167673111
        98 4.32743768692017
        99 4.29282891750336
        100 4.26445244550705
        101 4.24075348377228
        102 4.21442580223083
        103 4.18931337594986
        104 4.16554138660431
        105 4.16620458364487
        106 4.13843885660171
        107 4.12009397745132
        108 4.10405675172806
        109 4.08187063932419
        110 4.06158328056335
        111 4.04238640069962
        112 4.02365955114365
        113 4.00631139278412
        114 3.9987352013588
        115 3.97893220186234
        116 3.96351670026779
        117 3.94817444086075
        118 3.93369679450989
        119 3.91975202560425
        120 3.90659111738205
        121 3.89394385814667
        122 3.88148363828659
        123 3.86983107328415
        124 3.85864186286926
        125 3.84797306060791
        126 3.83354594707489
        127 3.82390245199203
        128 3.81389416456223
        129 3.80463848114014
        130 3.79595324993134
        131 3.78766919374466
        132 3.77977210283279
        133 3.77223242521286
        134 3.76501324176788
        135 3.75807917118073
        136 3.75146416425705
        137 3.74632339477539
        138 3.74114272594452
        139 3.76992295980453
        140 3.74594091176987
        141 3.73699102401733
        142 3.73016721010208
        143 3.72251478433609
        144 3.71641288995743
        145 3.710551404953
        146 3.70586644411087
        147 3.70109314918518
        148 3.69571813344955
        149 3.69190980195999
        150 3.68731976747513
        151 3.68287785053253
        152 3.67839407920837
        153 3.67395573854446
        154 3.66958630084991
        155 3.66532958745956
        156 3.67959176301956
        157 3.672585272789
        158 3.66602780818939
        159 3.65938845872879
        160 3.65319255590439
        161 3.64764298200607
        162 3.64267330169678
        163 3.6383109331131
        164 3.63447645902634
        165 3.63091032505035
        166 3.62580258846283
        167 3.62230269908905
        168 3.61877599954605
        169 3.62003062963486
        170 3.61393266916275
        171 3.60803512334824
        172 3.60378994941711
        173 3.60064904689789
        174 3.59759813547134
        175 3.59469577074051
        176 3.59319584369659
        177 3.58990358114243
        178 3.58657890558243
        179 3.58326998949051
        180 3.59117008447647
      };
    \addplot [thick, darkorange25512714]
    table {%
        60 6.71324036121368
        61 6.45023798942566
        62 6.18862941265106
        63 6.0119056224823
        64 5.76102278232574
        65 5.51663448810577
        66 5.32611773014069
        67 5.19511370658874
        68 5.09217231273651
        69 4.96359603404999
        70 4.87179129123688
        71 4.9325309753418
        72 4.78883309364319
        73 4.7780361533165
        74 4.54916459321976
        75 4.47003574371338
        76 4.38462271690369
        77 4.31286084651947
        78 4.35359139442444
        79 4.28676338195801
        80 4.22873116731644
        81 4.17930256128311
        82 4.14619889259338
        83 4.11437313556671
        84 4.08262405395508
        85 4.059011054039
        86 4.03390507698059
        87 4.01117920875549
        88 3.98938623666763
        89 3.96816674470901
        90 3.94787442684174
        91 3.92829800844193
        92 3.90930950641632
        93 3.89169080257416
        94 3.87473846673965
        95 3.85876148939133
        96 3.84545220136642
        97 3.86402628421783
        98 3.84585319757462
        99 3.82481114864349
        100 3.80665273666382
        101 3.7903829574585
        102 3.77604957818985
        103 3.76404137611389
        104 3.75866371393204
        105 3.74883054494858
        106 3.76520229578018
        107 3.73656487464905
        108 3.73039308786392
        109 3.71811655759811
        110 3.70962142944336
        111 3.70122314691544
        112 3.69375072717667
        113 3.68668297529221
        114 3.67949525117874
        115 3.6770535826683
        116 3.66937363147736
        117 3.65720311403275
        118 3.64831365346909
        119 3.63614109754562
        120 3.6278129696846
        121 3.63052182197571
        122 3.62138447761536
        123 3.61395283937454
        124 3.60629014968872
        125 3.59943286180496
        126 3.59327231645584
        127 3.58764562606812
        128 3.60516700744629
        129 3.63606952428818
        130 3.62098288536072
        131 3.60385907888412
        132 3.58736461400986
        133 3.57323348522186
        134 3.56211814880371
        135 3.55409859418869
        136 3.54865875244141
        137 3.54501187801361
        138 3.56483137607574
        139 3.55673438310623
        140 3.55464388132095
        141 3.54643287658691
        142 3.53922334909439
        143 3.59299912452698
        144 3.58435211181641
        145 3.58168736696243
        146 3.5349418759346
        147 3.52616338729858
        148 3.51773422956467
        149 3.51185252666473
        150 3.50693162679672
        151 3.50332306623459
        152 3.50028874874115
        153 3.49762337207794
        154 3.49514470100403
        155 3.49268684387207
        156 3.4901624083519
        157 3.48753625154495
        158 3.48480589389801
        159 3.48199793100357
        160 3.47915195226669
        161 3.48674125671387
        162 3.48256344795227
        163 3.47775501012802
        164 3.47294306755066
        165 3.46839860677719
        166 3.46432031393051
        167 3.46077462434769
        168 3.45725530385971
        169 3.45482754707336
        170 3.4523201584816
        171 3.44997133016586
        172 3.44762305021286
        173 3.4452082157135
        174 3.44270294904709
        175 3.44012115001678
        176 3.43749103546143
        177 3.43508951663971
        178 3.43084545135498
        179 3.42839332818985
        180 3.4255476474762
      };

  \end{axis}

\end{tikzpicture}

%% file: figs/pgfplot/ICCAD-L_mean_std/ICCAD-L_mean_std.tex
\pgfplotsset{
  width=\linewidth,
  height=.5\linewidth
}
\noindent
\begin{tikzpicture}[inner sep=2pt, outer sep=0pt]
\definecolor{darkgray176}{RGB}{176,176,176}

\definecolor{forestgreen4416044}{RGB}{44,160,44}
\definecolor{lightgray204}{RGB}{204,204,204}
\definecolor{steelblue31119180}{RGB}{31,119,180}
\definecolor{crimson2143940}{RGB}{214,39,40}
\definecolor{mediumpurple148103189}{RGB}{148,103,189}
\definecolor{darkorange25512714}{RGB}{255,127,14}

\begin{axis}[
legend cell align={left},
legend style={
    fill opacity=0.8,
    draw opacity=1,
    text opacity=1,
    draw=none,
    font=\huge
    },
xmin=54, xmax=186,
ymin=6.07970236559904, ymax=22.5294088052295,
tick align=inside,
tick pos=left,
tick style={major tick length=3pt},
xtick = {60, 90, 150, 180},
xticklabels = {20, 30, 50, 60},
ytick = {9, 18, 21},
yticklabels = {3, 6, 7},
xlabel = {step},
xlabel near ticks,
xlabel shift={-8pt},
ylabel= {$\mathcal{L}_{smo}$},
ylabel near ticks,
ylabel shift={-5pt},
]
\path [draw=crimson2143940, fill=crimson2143940, opacity=0.4]
(axis cs:60,18.6071732273002)
--(axis cs:60,15.0703284511666)
--(axis cs:61,14.4255764407211)
--(axis cs:62,13.7559589984732)
--(axis cs:63,13.0761353612652)
--(axis cs:64,12.4076503377283)
--(axis cs:65,11.7790021403111)
--(axis cs:66,11.2196839105897)
--(axis cs:67,10.751515509692)
--(axis cs:68,10.3811398771371)
--(axis cs:69,10.1001031922388)
--(axis cs:70,9.89230801144666)
--(axis cs:71,9.73230625480906)
--(axis cs:72,9.58320764018641)
--(axis cs:73,9.4133094970538)
--(axis cs:74,9.21427031394844)
--(axis cs:75,8.99880826147696)
--(axis cs:76,8.78806694905425)
--(axis cs:77,8.60195608262903)
--(axis cs:78,8.4531920757053)
--(axis cs:79,8.34441152463441)
--(axis cs:80,8.26963239544467)
--(axis cs:81,8.21819692347408)
--(axis cs:82,8.17872685711259)
--(axis cs:83,8.14189760092502)
--(axis cs:84,8.10182649397162)
--(axis cs:85,8.0562871807419)
--(axis cs:86,8.00617064338728)
--(axis cs:87,7.95448476075616)
--(axis cs:88,7.90513473068624)
--(axis cs:89,7.8616595636728)
--(axis cs:90,7.82619638315012)
--(axis cs:91,7.79890193819546)
--(axis cs:92,7.77797813594992)
--(axis cs:93,7.76039006014989)
--(axis cs:94,7.74302004434343)
--(axis cs:95,7.72381826272915)
--(axis cs:96,7.7024092485331)
--(axis cs:97,7.67991083685532)
--(axis cs:98,7.65811869382479)
--(axis cs:99,7.63860638287939)
--(axis cs:100,7.62214417029074)
--(axis cs:101,7.60861489350486)
--(axis cs:102,7.5972703860222)
--(axis cs:103,7.58714539985419)
--(axis cs:104,7.57739889635679)
--(axis cs:105,7.56751321671574)
--(axis cs:106,7.557330703199)
--(axis cs:107,7.54699027948905)
--(axis cs:108,7.53679979401921)
--(axis cs:109,7.52709089609689)
--(axis cs:110,7.51810936436526)
--(axis cs:111,7.50995174766106)
--(axis cs:112,7.50255210393221)
--(axis cs:113,7.49573350150322)
--(axis cs:114,7.4892833322525)
--(axis cs:115,7.48301634184256)
--(axis cs:116,7.47681870242928)
--(axis cs:117,7.47065962713652)
--(axis cs:118,7.46457434526943)
--(axis cs:119,7.45862712512304)
--(axis cs:120,7.4528815923192)
--(axis cs:121,7.4473790993748)
--(axis cs:122,7.442127869692)
--(axis cs:123,7.43711218408217)
--(axis cs:124,7.43230135142663)
--(axis cs:125,7.42766437819053)
--(axis cs:126,7.4231742822316)
--(axis cs:127,7.41881841519926)
--(axis cs:128,7.41458914332646)
--(axis cs:129,7.41048036113936)
--(axis cs:130,7.40648184748801)
--(axis cs:131,7.40256893851018)
--(axis cs:132,7.39871563487904)
--(axis cs:133,7.3949081383511)
--(axis cs:134,7.39115028966358)
--(axis cs:135,7.38745650041494)
--(axis cs:136,7.38384600921644)
--(axis cs:137,7.3803214573631)
--(axis cs:138,7.37689135074152)
--(axis cs:139,7.37355307961186)
--(axis cs:140,7.37030427080796)
--(axis cs:141,7.36714358383838)
--(axis cs:142,7.36406258672366)
--(axis cs:143,7.36105533616899)
--(axis cs:144,7.35810805501553)
--(axis cs:145,7.35521332624091)
--(axis cs:146,7.35235627441237)
--(axis cs:147,7.34953076794762)
--(axis cs:148,7.3467293179419)
--(axis cs:149,7.34394961983883)
--(axis cs:150,7.34119301783529)
--(axis cs:151,7.33846542259088)
--(axis cs:152,7.33576609851679)
--(axis cs:153,7.33309841948686)
--(axis cs:154,7.33045894358037)
--(axis cs:155,7.32784345956313)
--(axis cs:156,7.32524701719004)
--(axis cs:157,7.32267096419829)
--(axis cs:158,7.32011156590509)
--(axis cs:159,7.31757122966019)
--(axis cs:160,7.31505265003589)
--(axis cs:161,7.31255675682524)
--(axis cs:162,7.31009179685105)
--(axis cs:163,7.30767047780112)
--(axis cs:164,7.30529649381555)
--(axis cs:165,7.30297960865396)
--(axis cs:166,7.30072359077299)
--(axis cs:167,7.29853352964284)
--(axis cs:168,7.2964080237656)
--(axis cs:169,7.29434635368392)
--(axis cs:170,7.2923430852048)
--(axis cs:171,7.29039367985433)
--(axis cs:172,7.2884913399838)
--(axis cs:173,7.28662923484952)
--(axis cs:174,7.2848046630918)
--(axis cs:175,7.28301267824437)
--(axis cs:176,7.28124996198471)
--(axis cs:177,7.27951250814057)
--(axis cs:178,7.27779616967335)
--(axis cs:179,7.27609275887576)
--(axis cs:180,7.27439889153749)
--(axis cs:180,10.4187499121734)
--(axis cs:180,10.4187499121734)
--(axis cs:179,10.4210248558703)
--(axis cs:178,10.4233358417402)
--(axis cs:177,10.4256857798233)
--(axis cs:176,10.4280763052577)
--(axis cs:175,10.430506608865)
--(axis cs:174,10.4329716354311)
--(axis cs:173,10.4354676851639)
--(axis cs:172,10.4379951521732)
--(axis cs:171,10.4405600135356)
--(axis cs:170,10.4431785174258)
--(axis cs:169,10.445872762527)
--(axis cs:168,10.4486638642044)
--(axis cs:167,10.4515650802815)
--(axis cs:166,10.4545784569565)
--(axis cs:165,10.4576944864999)
--(axis cs:164,10.4608868824967)
--(axis cs:163,10.4641264162724)
--(axis cs:162,10.4673835602046)
--(axis cs:161,10.4706561910202)
--(axis cs:160,10.4739697002754)
--(axis cs:159,10.4773452093962)
--(axis cs:158,10.4807794520087)
--(axis cs:157,10.4842475614975)
--(axis cs:156,10.4877236210851)
--(axis cs:155,10.4911955991222)
--(axis cs:154,10.494660456444)
--(axis cs:153,10.4981295506364)
--(axis cs:152,10.5016139292542)
--(axis cs:151,10.5051374254621)
--(axis cs:150,10.5087062741569)
--(axis cs:149,10.5123372873572)
--(axis cs:148,10.5160387599084)
--(axis cs:147,10.5198171858488)
--(axis cs:146,10.5236735984819)
--(axis cs:145,10.5276121437298)
--(axis cs:144,10.5316270714798)
--(axis cs:143,10.5357138059819)
--(axis cs:142,10.5398679819714)
--(axis cs:141,10.5440781111079)
--(axis cs:140,10.5483354844411)
--(axis cs:139,10.5526393947534)
--(axis cs:138,10.5569938421296)
--(axis cs:137,10.5614159083596)
--(axis cs:136,10.5659405584716)
--(axis cs:135,10.5705959440431)
--(axis cs:134,10.5754126490266)
--(axis cs:133,10.5804026641086)
--(axis cs:132,10.5855712456141)
--(axis cs:131,10.5909176566436)
--(axis cs:130,10.5964375403294)
--(axis cs:129,10.6021199653987)
--(axis cs:128,10.6079581016992)
--(axis cs:127,10.6139383902207)
--(axis cs:126,10.6200627363536)
--(axis cs:125,10.6263376207108)
--(axis cs:124,10.6327773839249)
--(axis cs:123,10.6394064468993)
--(axis cs:122,10.6462409276102)
--(axis cs:121,10.6532870421352)
--(axis cs:120,10.6605415644191)
--(axis cs:119,10.6679891621084)
--(axis cs:118,10.6756074822757)
--(axis cs:117,10.6833731754453)
--(axis cs:116,10.6912665865722)
--(axis cs:115,10.6992838443147)
--(axis cs:114,10.7074603467941)
--(axis cs:113,10.7158656577375)
--(axis cs:112,10.7246076822922)
--(axis cs:111,10.7338236200758)
--(axis cs:110,10.7436466743005)
--(axis cs:109,10.7541701665252)
--(axis cs:108,10.7654165450921)
--(axis cs:107,10.7773105818696)
--(axis cs:106,10.7897188192009)
--(axis cs:105,10.8024681675616)
--(axis cs:104,10.8154194592035)
--(axis cs:103,10.8285204174066)
--(axis cs:102,10.841913230616)
--(axis cs:101,10.855972318104)
--(axis cs:100,10.8712661309178)
--(axis cs:99,10.8884299978884)
--(axis cs:98,10.9079917264023)
--(axis cs:97,10.9301774161659)
--(axis cs:96,10.9547887037374)
--(axis cs:95,10.9812218106656)
--(axis cs:94,11.0086482181383)
--(axis cs:93,11.0363841364749)
--(axis cs:92,11.0643146496946)
--(axis cs:91,11.0932034504459)
--(axis cs:90,11.1246067059917)
--(axis cs:89,11.1603411306021)
--(axis cs:88,11.2017610975608)
--(axis cs:87,11.2490914225629)
--(axis cs:86,11.3011235346218)
--(axis cs:85,11.3553340083755)
--(axis cs:84,11.4084154913494)
--(axis cs:83,11.4575863659024)
--(axis cs:82,11.5020281859172)
--(axis cs:81,11.5439422061646)
--(axis cs:80,11.5890639699118)
--(axis cs:79,11.6463861857843)
--(axis cs:78,11.726841703439)
--(axis cs:77,11.8407593714725)
--(axis cs:76,11.9944775684628)
--(axis cs:75,12.1871485027442)
--(axis cs:74,12.4093880570232)
--(axis cs:73,12.6449655349897)
--(axis cs:72,12.8756252913321)
--(axis cs:71,13.0900553098272)
--(axis cs:70,13.2950365396398)
--(axis cs:69,13.5172958327246)
--(axis cs:68,13.7878767702018)
--(axis cs:67,14.1281492928592)
--(axis cs:66,14.5526744115539)
--(axis cs:65,15.0703512685024)
--(axis cs:64,15.6786011118567)
--(axis cs:63,16.3611239313373)
--(axis cs:62,17.0941661235971)
--(axis cs:61,17.851377828688)
--(axis cs:60,18.6071732273002)
--cycle;

\path [draw=mediumpurple148103189, fill=mediumpurple148103189, opacity=0.4]
(axis cs:60,22.163670983415)
--(axis cs:60,13.7601966237016)
--(axis cs:61,13.4765331828552)
--(axis cs:62,13.21315350604)
--(axis cs:63,12.301620654888)
--(axis cs:64,11.7863062680259)
--(axis cs:65,11.305431531719)
--(axis cs:66,10.9717833429924)
--(axis cs:67,10.471227727761)
--(axis cs:68,10.0181830957245)
--(axis cs:69,9.66605416350895)
--(axis cs:70,9.37148671590041)
--(axis cs:71,9.36111975351397)
--(axis cs:72,9.31237591043469)
--(axis cs:73,9.26367267941655)
--(axis cs:74,9.24686428453423)
--(axis cs:75,9.16643810670815)
--(axis cs:76,9.05909567199239)
--(axis cs:77,8.8910685341334)
--(axis cs:78,8.70654621230182)
--(axis cs:79,8.52471878523028)
--(axis cs:80,8.36077163175634)
--(axis cs:81,8.23642754482652)
--(axis cs:82,8.14897587981633)
--(axis cs:83,8.08741506120489)
--(axis cs:84,8.02891270161136)
--(axis cs:85,7.99117014711477)
--(axis cs:86,7.95815880451357)
--(axis cs:87,7.9183665755601)
--(axis cs:88,7.86717821419247)
--(axis cs:89,7.81858454102713)
--(axis cs:90,7.76509787790817)
--(axis cs:91,7.70827813182022)
--(axis cs:92,7.65120667002635)
--(axis cs:93,7.59789330824127)
--(axis cs:94,7.55007949792904)
--(axis cs:95,7.50747979502794)
--(axis cs:96,7.46953383497379)
--(axis cs:97,7.43458518156524)
--(axis cs:98,7.40183539222629)
--(axis cs:99,7.36845910490082)
--(axis cs:100,7.33730283467115)
--(axis cs:101,7.30556341103548)
--(axis cs:102,7.27552936689177)
--(axis cs:103,7.2488636052468)
--(axis cs:104,7.22446737758295)
--(axis cs:105,7.20128658274087)
--(axis cs:106,7.10466652374883)
--(axis cs:107,7.09350144565427)
--(axis cs:108,7.07388290559593)
--(axis cs:109,7.06530945286133)
--(axis cs:110,7.04695091740346)
--(axis cs:111,7.0312920499835)
--(axis cs:112,7.01588440789427)
--(axis cs:113,7.00028614597671)
--(axis cs:114,6.98478679018001)
--(axis cs:115,6.96950063620761)
--(axis cs:116,6.95482648864994)
--(axis cs:117,6.95329630215775)
--(axis cs:118,6.94066256696508)
--(axis cs:119,6.9213015292074)
--(axis cs:120,6.91064882387811)
--(axis cs:121,6.90507772224165)
--(axis cs:122,6.90030790970918)
--(axis cs:123,6.89716853874815)
--(axis cs:124,6.89261259979246)
--(axis cs:125,6.88842321764195)
--(axis cs:126,6.89536430617176)
--(axis cs:127,6.8767204798387)
--(axis cs:128,6.86979800480463)
--(axis cs:129,6.86216950853023)
--(axis cs:130,6.85424777273725)
--(axis cs:131,6.84674048234303)
--(axis cs:132,6.83964802373075)
--(axis cs:133,6.88368458737496)
--(axis cs:134,6.8732607383397)
--(axis cs:135,6.85991783845736)
--(axis cs:136,6.83865305506505)
--(axis cs:137,6.82535508630317)
--(axis cs:138,6.81565965125094)
--(axis cs:139,6.80916901558095)
--(axis cs:140,6.80318297559172)
--(axis cs:141,6.79847055501886)
--(axis cs:142,6.76755364668431)
--(axis cs:143,6.75875043155761)
--(axis cs:144,6.77058229199028)
--(axis cs:145,6.7721952574614)
--(axis cs:146,6.7697686362015)
--(axis cs:147,6.76643526394797)
--(axis cs:148,6.76250552889231)
--(axis cs:149,6.75820139178625)
--(axis cs:150,6.75079626661225)
--(axis cs:151,6.74611755908186)
--(axis cs:152,6.74163448514329)
--(axis cs:153,6.73804275212601)
--(axis cs:154,6.73315589259457)
--(axis cs:155,6.72869894456637)
--(axis cs:156,6.7244765814304)
--(axis cs:157,6.72044257581948)
--(axis cs:158,6.71661423488737)
--(axis cs:159,6.7126955517207)
--(axis cs:160,6.70878520617929)
--(axis cs:161,6.70514101891551)
--(axis cs:162,6.70618640255413)
--(axis cs:163,6.70075235290539)
--(axis cs:164,6.69613303854216)
--(axis cs:165,6.69178804577916)
--(axis cs:166,6.68728560744731)
--(axis cs:167,6.6836140762657)
--(axis cs:168,6.68028315316022)
--(axis cs:169,6.67725758279331)
--(axis cs:170,6.67443553000106)
--(axis cs:171,6.671705141374)
--(axis cs:172,6.66897621585751)
--(axis cs:173,6.66619583037507)
--(axis cs:174,6.66333934329031)
--(axis cs:175,6.66038864434258)
--(axis cs:176,6.6573283957104)
--(axis cs:177,6.65435850467582)
--(axis cs:178,6.6514193522956)
--(axis cs:179,6.64851613013856)
--(axis cs:180,6.64500129267074)
--(axis cs:180,10.6581313367238)
--(axis cs:180,10.6581313367238)
--(axis cs:179,10.663014936756)
--(axis cs:178,10.6683704388116)
--(axis cs:177,10.6739253774653)
--(axis cs:176,10.6796663476368)
--(axis cs:175,10.6856280487727)
--(axis cs:174,10.6918567512226)
--(axis cs:173,10.698343697518)
--(axis cs:172,10.705093951684)
--(axis cs:171,10.7121666046892)
--(axis cs:170,10.7196644684731)
--(axis cs:169,10.7277349976205)
--(axis cs:168,10.7365464328688)
--(axis cs:167,10.7462166656166)
--(axis cs:166,10.7567726200917)
--(axis cs:165,10.7681869107333)
--(axis cs:164,10.7797287969376)
--(axis cs:163,10.7919016845702)
--(axis cs:162,10.8046769969515)
--(axis cs:161,10.8114209661004)
--(axis cs:160,10.8221109806493)
--(axis cs:159,10.8316320888127)
--(axis cs:158,10.8413283462894)
--(axis cs:157,10.852744489305)
--(axis cs:156,10.8634643975735)
--(axis cs:155,10.87432296324)
--(axis cs:154,10.8854104297321)
--(axis cs:153,10.8970011550395)
--(axis cs:152,10.908206728083)
--(axis cs:151,10.9225469916994)
--(axis cs:150,10.9392309703262)
--(axis cs:149,10.9728841947616)
--(axis cs:148,10.9987338090288)
--(axis cs:147,11.0286634222703)
--(axis cs:146,11.0650313210739)
--(axis cs:145,11.1014829499361)
--(axis cs:144,11.135881426424)
--(axis cs:143,11.1521726679602)
--(axis cs:142,11.1506404565567)
--(axis cs:141,11.0344882386213)
--(axis cs:140,11.0562240678653)
--(axis cs:139,11.0781793502027)
--(axis cs:138,11.1079611449527)
--(axis cs:137,11.1276432657476)
--(axis cs:136,11.1410307256051)
--(axis cs:135,11.1233291555612)
--(axis cs:134,11.1484760742518)
--(axis cs:133,11.1655759335601)
--(axis cs:132,11.1898799265561)
--(axis cs:131,11.2050079841487)
--(axis cs:130,11.2211513399642)
--(axis cs:129,11.2405877546534)
--(axis cs:128,11.2621282456531)
--(axis cs:127,11.2829090558363)
--(axis cs:126,11.3871965191761)
--(axis cs:125,11.3144813882235)
--(axis cs:124,11.3401402955628)
--(axis cs:123,11.3747488997589)
--(axis cs:122,11.4145662625015)
--(axis cs:121,11.4852326606014)
--(axis cs:120,11.4424151409656)
--(axis cs:119,11.4627710606669)
--(axis cs:118,11.5524649698086)
--(axis cs:117,11.5600824133574)
--(axis cs:116,11.4801221464514)
--(axis cs:115,11.4960807808761)
--(axis cs:114,11.5139991824152)
--(axis cs:113,11.5348375837291)
--(axis cs:112,11.5588570509985)
--(axis cs:111,11.5864844392743)
--(axis cs:110,11.6170130680397)
--(axis cs:109,11.6475456954541)
--(axis cs:108,11.6943808158701)
--(axis cs:107,11.7234467774597)
--(axis cs:106,11.7810219718872)
--(axis cs:105,11.5774153902206)
--(axis cs:104,11.6082287741504)
--(axis cs:103,11.6406832659385)
--(axis cs:102,11.675801104147)
--(axis cs:101,11.7149962813473)
--(axis cs:100,11.7514645984954)
--(axis cs:99,11.7864607101627)
--(axis cs:98,11.8247432248697)
--(axis cs:97,11.8630409323264)
--(axis cs:96,11.9012061018548)
--(axis cs:95,11.9417924561012)
--(axis cs:94,11.9880354312702)
--(axis cs:93,12.0421235336472)
--(axis cs:92,12.1028004787163)
--(axis cs:91,12.1729600426183)
--(axis cs:90,12.2482513690706)
--(axis cs:89,12.3306113207893)
--(axis cs:88,12.4170962685346)
--(axis cs:87,12.4801840302138)
--(axis cs:86,12.5880661710724)
--(axis cs:85,12.6991100042524)
--(axis cs:84,12.8228895521118)
--(axis cs:83,12.9756312034436)
--(axis cs:82,13.1400812796075)
--(axis cs:81,13.3229033954201)
--(axis cs:80,13.5264964346499)
--(axis cs:79,13.742054724749)
--(axis cs:78,13.9924629200834)
--(axis cs:77,14.2573849875951)
--(axis cs:76,14.5184910742234)
--(axis cs:75,14.7198837240414)
--(axis cs:74,15.0114858970454)
--(axis cs:73,15.1891238751489)
--(axis cs:72,15.4600497411156)
--(axis cs:71,15.9078778203386)
--(axis cs:70,16.2952474073296)
--(axis cs:69,16.9102645773358)
--(axis cs:68,17.5072369024444)
--(axis cs:67,18.3217679158548)
--(axis cs:66,19.118598470057)
--(axis cs:65,19.3700376760447)
--(axis cs:64,20.0374640643106)
--(axis cs:63,20.6468992425729)
--(axis cs:62,20.9730250828272)
--(axis cs:61,21.690718213324)
--(axis cs:60,22.163670983415)
--cycle;

\path [draw=darkorange25512714, fill=darkorange25512714, opacity=0.4]
(axis cs:60,20.1151480732353)
--(axis cs:60,12.875398344001)
--(axis cs:61,12.3488489361872)
--(axis cs:62,11.8105194805137)
--(axis cs:63,11.2142918415285)
--(axis cs:64,10.8110019844116)
--(axis cs:65,10.4106556891826)
--(axis cs:66,10.1037535337893)
--(axis cs:67,9.8592554135496)
--(axis cs:68,9.64290890101594)
--(axis cs:69,9.50567070330865)
--(axis cs:70,9.34961407933514)
--(axis cs:71,9.19931385474199)
--(axis cs:72,9.07248176325827)
--(axis cs:73,8.64228709569441)
--(axis cs:74,8.52734083529531)
--(axis cs:75,8.44829138405463)
--(axis cs:76,8.40228508064599)
--(axis cs:77,8.28671032771244)
--(axis cs:78,8.19081373181896)
--(axis cs:79,8.12095746325107)
--(axis cs:80,8.06545075360748)
--(axis cs:81,8.01061942038783)
--(axis cs:82,7.96714092052201)
--(axis cs:83,7.92521865379131)
--(axis cs:84,7.90529139650455)
--(axis cs:85,7.85938246688622)
--(axis cs:86,7.81071882977708)
--(axis cs:87,7.76373863588389)
--(axis cs:88,7.72126633662396)
--(axis cs:89,7.68383151053481)
--(axis cs:90,7.65130406531303)
--(axis cs:91,7.62099665887897)
--(axis cs:92,7.597544386758)
--(axis cs:93,7.57459654950842)
--(axis cs:94,7.55034703449196)
--(axis cs:95,7.53586088772675)
--(axis cs:96,7.51133736894472)
--(axis cs:97,7.48548663944557)
--(axis cs:98,7.46159119544644)
--(axis cs:99,7.44073984860956)
--(axis cs:100,7.42228787007429)
--(axis cs:101,7.40701256854681)
--(axis cs:102,7.39344424847251)
--(axis cs:103,7.38112738747045)
--(axis cs:104,7.36913460529263)
--(axis cs:105,7.35838327169668)
--(axis cs:106,7.44529193118886)
--(axis cs:107,7.42018437143753)
--(axis cs:108,7.39239296262415)
--(axis cs:109,7.36425628361836)
--(axis cs:110,7.33884145055993)
--(axis cs:111,7.3174967335888)
--(axis cs:112,7.30017284442372)
--(axis cs:113,7.28623652479514)
--(axis cs:114,7.27472899628539)
--(axis cs:115,7.26469060878624)
--(axis cs:116,7.25537054585823)
--(axis cs:117,7.24629191673443)
--(axis cs:118,7.23595776968086)
--(axis cs:119,7.22351310540372)
--(axis cs:120,7.224026307596)
--(axis cs:121,7.21366068285895)
--(axis cs:122,7.20505322701187)
--(axis cs:123,7.1950830039346)
--(axis cs:124,7.18588934669729)
--(axis cs:125,7.17770361393466)
--(axis cs:126,7.17006913139602)
--(axis cs:127,7.16224835522794)
--(axis cs:128,7.15446473388889)
--(axis cs:129,7.14640949016066)
--(axis cs:130,7.14048695499091)
--(axis cs:131,7.13450593347309)
--(axis cs:132,7.12841574852633)
--(axis cs:133,7.12231567955224)
--(axis cs:134,7.11630119381155)
--(axis cs:135,7.14541799485202)
--(axis cs:136,7.13786354801498)
--(axis cs:137,7.10943290925157)
--(axis cs:138,7.09718022042389)
--(axis cs:139,7.08752811775707)
--(axis cs:140,7.07989825903453)
--(axis cs:141,7.07487023823572)
--(axis cs:142,7.0710981039517)
--(axis cs:143,7.06777900798944)
--(axis cs:144,7.06445839851256)
--(axis cs:145,7.06058499740007)
--(axis cs:146,7.05603480895991)
--(axis cs:147,7.05093243660916)
--(axis cs:148,7.04554707165494)
--(axis cs:149,7.04018668113043)
--(axis cs:150,7.03517391586249)
--(axis cs:151,7.03059818130372)
--(axis cs:152,7.02906397406926)
--(axis cs:153,7.03118199558531)
--(axis cs:154,7.02732793365428)
--(axis cs:155,7.01778111163661)
--(axis cs:156,7.01895625131312)
--(axis cs:157,7.01592546037083)
--(axis cs:158,7.00683230968059)
--(axis cs:159,7.0015634342316)
--(axis cs:160,6.99641339910494)
--(axis cs:161,6.99237260113168)
--(axis cs:162,7.02288426788541)
--(axis cs:163,7.01395580834361)
--(axis cs:164,7.00525548872165)
--(axis cs:165,6.99617409708731)
--(axis cs:166,6.98781366959077)
--(axis cs:167,6.98054051289849)
--(axis cs:168,6.97572903519049)
--(axis cs:169,6.97216184019299)
--(axis cs:170,6.9690623171926)
--(axis cs:171,6.96563234650532)
--(axis cs:172,6.96149869551153)
--(axis cs:173,6.95673778445941)
--(axis cs:174,6.95169281677922)
--(axis cs:175,6.94685849551153)
--(axis cs:176,6.94266871681798)
--(axis cs:177,6.93905919965057)
--(axis cs:178,6.93596950205236)
--(axis cs:179,6.93317599985)
--(axis cs:180,7.07382804504116)
--(axis cs:180,10.2329868162565)
--(axis cs:180,10.2329868162565)
--(axis cs:179,10.2429665019811)
--(axis cs:178,10.2465842661056)
--(axis cs:177,10.2498196226761)
--(axis cs:176,10.2530478549899)
--(axis cs:175,10.2565450727563)
--(axis cs:174,10.2614237336663)
--(axis cs:173,10.267515965388)
--(axis cs:172,10.2746515692285)
--(axis cs:171,10.2828575102146)
--(axis cs:170,10.2918265454173)
--(axis cs:169,10.3010424864366)
--(axis cs:168,10.3099229347192)
--(axis cs:167,10.3178080092878)
--(axis cs:166,10.3240873275711)
--(axis cs:165,10.3282540797926)
--(axis cs:164,10.3323619085867)
--(axis cs:163,10.3347090476134)
--(axis cs:162,10.3409524020365)
--(axis cs:161,10.3577502321203)
--(axis cs:160,10.349402831669)
--(axis cs:159,10.3540680126068)
--(axis cs:158,10.3654104271358)
--(axis cs:157,10.4380062527334)
--(axis cs:156,10.513597993682)
--(axis cs:155,10.5864251643033)
--(axis cs:154,10.4532890173345)
--(axis cs:153,10.4831803396198)
--(axis cs:152,10.3965672629608)
--(axis cs:151,10.3830236257947)
--(axis cs:150,10.3870683441168)
--(axis cs:149,10.3917397600622)
--(axis cs:148,10.3972357118534)
--(axis cs:147,10.403680199957)
--(axis cs:146,10.4110413972474)
--(axis cs:145,10.4190148599303)
--(axis cs:144,10.4273413375104)
--(axis cs:143,10.4357846821961)
--(axis cs:142,10.4223017068393)
--(axis cs:141,10.4284810686796)
--(axis cs:140,10.434834483275)
--(axis cs:139,10.4414018787334)
--(axis cs:138,10.4484578639877)
--(axis cs:137,10.4561353471266)
--(axis cs:136,10.4651385233656)
--(axis cs:135,10.4740304762173)
--(axis cs:134,10.4821638673762)
--(axis cs:133,10.4915724887827)
--(axis cs:132,10.5011795311673)
--(axis cs:131,10.511478239348)
--(axis cs:130,10.5201769358477)
--(axis cs:129,10.5287377571538)
--(axis cs:128,10.5392433998086)
--(axis cs:127,10.5486903845487)
--(axis cs:126,10.5599803833745)
--(axis cs:125,10.5744139245175)
--(axis cs:124,10.5919783519721)
--(axis cs:123,10.6135096016562)
--(axis cs:122,10.6427620958641)
--(axis cs:121,10.6595019682031)
--(axis cs:120,10.6852404503691)
--(axis cs:119,10.7112954349405)
--(axis cs:118,10.6431250054351)
--(axis cs:117,10.6619197913344)
--(axis cs:116,10.674318766123)
--(axis cs:115,10.6860317995084)
--(axis cs:114,10.6981766395484)
--(axis cs:113,10.7131706712498)
--(axis cs:112,10.7361520380836)
--(axis cs:111,10.7716091585926)
--(axis cs:110,10.7987450477006)
--(axis cs:109,10.8151945144163)
--(axis cs:108,10.8027736252054)
--(axis cs:107,10.8238401913982)
--(axis cs:106,10.8322468452089)
--(axis cs:105,10.8688819909071)
--(axis cs:104,10.8737626191336)
--(axis cs:103,10.8923870740469)
--(axis cs:102,10.9125151669633)
--(axis cs:101,10.9351229562029)
--(axis cs:100,10.9580367797365)
--(axis cs:99,10.9857882237476)
--(axis cs:98,11.0153691774784)
--(axis cs:97,11.0481586089675)
--(axis cs:96,11.0901843137945)
--(axis cs:95,11.1244718873892)
--(axis cs:94,11.159696300469)
--(axis cs:93,11.2093771443392)
--(axis cs:92,11.2459784480198)
--(axis cs:91,11.2872237276262)
--(axis cs:90,11.3386488803962)
--(axis cs:89,11.3891501331419)
--(axis cs:88,11.4409991548616)
--(axis cs:87,11.4921784364123)
--(axis cs:86,11.5417600558736)
--(axis cs:85,11.5979363159202)
--(axis cs:84,11.647810524577)
--(axis cs:83,11.644041561968)
--(axis cs:82,11.6851265355327)
--(axis cs:81,11.7423199945902)
--(axis cs:80,11.8189655404904)
--(axis cs:79,11.9344766683712)
--(axis cs:78,12.0940012935106)
--(axis cs:77,12.3053467096411)
--(axis cs:76,12.5963513558164)
--(axis cs:75,13.3035995613799)
--(axis cs:74,13.8435182154554)
--(axis cs:73,14.205971971841)
--(axis cs:72,13.3402433992478)
--(axis cs:71,13.4992609743596)
--(axis cs:70,13.773764674327)
--(axis cs:69,13.9695450178217)
--(axis cs:68,14.2072192728122)
--(axis cs:67,14.6366416888064)
--(axis cs:66,15.2212821336301)
--(axis cs:65,15.8714943886372)
--(axis cs:64,16.8421628791748)
--(axis cs:63,17.7973542384886)
--(axis cs:62,18.1821645023354)
--(axis cs:61,19.2168508318792)
--(axis cs:60,20.1151480732353)
--cycle;

\addplot [thick, crimson2143940]
table {%
60 16.8387508392334
61 16.1384771347046
62 15.4250625610352
63 14.7186296463013
64 14.0431257247925
65 13.4246767044067
66 12.8861791610718
67 12.4398324012756
68 12.0845083236694
69 11.8086995124817
70 11.5936722755432
71 11.4111807823181
72 11.2294164657593
73 11.0291375160217
74 10.8118291854858
75 10.5929783821106
76 10.3912722587585
77 10.2213577270508
78 10.0900168895721
79 9.99539885520935
80 9.92934818267822
81 9.88106956481934
82 9.84037752151489
83 9.7997419834137
84 9.75512099266052
85 9.70581059455872
86 9.65364708900452
87 9.60178809165955
88 9.55344791412353
89 9.51100034713745
90 9.47540154457092
91 9.44605269432068
92 9.42114639282227
93 9.39838709831238
94 9.37583413124085
95 9.35252003669739
96 9.32859897613525
97 9.30504412651062
98 9.28305521011353
99 9.26351819038391
100 9.24670515060425
101 9.23229360580444
102 9.21959180831909
103 9.20783290863037
104 9.19640917778015
105 9.18499069213867
106 9.17352476119995
107 9.16215043067932
108 9.15110816955566
109 9.14063053131104
110 9.13087801933289
111 9.12188768386841
112 9.11357989311218
113 9.10579957962036
114 9.09837183952332
115 9.09115009307861
116 9.08404264450073
117 9.07701640129089
118 9.07009091377258
119 9.06330814361572
120 9.05671157836914
121 9.05033307075501
122 9.04418439865112
123 9.03825931549072
124 9.03253936767578
125 9.02700099945068
126 9.0216185092926
127 9.01637840270996
128 9.01127362251282
129 9.00630016326904
130 9.00145969390869
131 8.9967432975769
132 8.99214344024658
133 8.98765540122986
134 8.98328146934509
135 8.979026222229
136 8.97489328384399
137 8.97086868286133
138 8.96694259643555
139 8.96309623718262
140 8.95931987762451
141 8.95561084747314
142 8.95196528434753
143 8.94838457107544
144 8.94486756324768
145 8.94141273498535
146 8.93801493644714
147 8.93467397689819
148 8.93138403892517
149 8.92814345359802
150 8.92494964599609
151 8.92180142402649
152 8.9186900138855
153 8.91561398506165
154 8.91255970001221
155 8.90951952934265
156 8.90648531913757
157 8.9034592628479
158 8.90044550895691
159 8.8974582195282
160 8.89451117515564
161 8.89160647392273
162 8.88873767852783
163 8.88589844703674
164 8.88309168815613
165 8.88033704757691
166 8.87765102386475
167 8.87504930496216
168 8.87253594398499
169 8.87010955810547
170 8.86776080131531
171 8.86547684669495
172 8.86324324607849
173 8.86104846000671
174 8.85888814926147
175 8.85675964355469
176 8.85466313362122
177 8.85259914398193
178 8.85056600570679
179 8.84855880737305
180 8.84657440185547
};
\addlegendentry{\texttt{BiSMO-FD} mean}
\addplot [thick, mediumpurple148103189]
table {%
60 17.9619338035583
61 17.5836256980896
62 17.0930892944336
63 16.4742599487305
64 15.9118851661682
65 15.3377346038818
66 15.0451909065247
67 14.3964978218079
68 13.7627099990845
69 13.2881593704224
70 12.833367061615
71 12.6344987869263
72 12.3862128257751
73 12.2263982772827
74 12.1291750907898
75 11.9431609153748
76 11.7887933731079
77 11.5742267608643
78 11.3495045661926
79 11.1333867549896
80 10.9436340332031
81 10.7796654701233
82 10.6445285797119
83 10.5315231323242
84 10.4259011268616
85 10.3451400756836
86 10.273112487793
87 10.199275302887
88 10.1421372413635
89 10.0745979309082
90 10.0066746234894
91 9.94061908721924
92 9.87700357437134
93 9.82000842094421
94 9.76905746459961
95 9.72463612556458
96 9.68536996841431
97 9.6488130569458
98 9.61328930854797
99 9.57745990753174
100 9.54438371658325
101 9.51027984619141
102 9.47566523551941
103 9.44477343559265
104 9.4163480758667
105 9.38935098648071
106 9.44284424781799
107 9.40847411155701
108 9.38413186073303
109 9.35642757415772
110 9.33198199272156
111 9.30888824462891
112 9.28737072944641
113 9.2675618648529
114 9.24939298629761
115 9.23279070854187
116 9.21747431755066
117 9.25668935775757
118 9.24656376838684
119 9.19203629493713
120 9.17653198242187
121 9.19515519142151
122 9.15743708610535
123 9.13595871925354
124 9.11637644767761
125 9.10145230293274
126 9.14128041267395
127 9.07981476783752
128 9.06596312522888
129 9.0513786315918
130 9.03769955635071
131 9.02587423324585
132 9.01476397514343
133 9.02463026046753
134 9.01086840629578
135 8.99162349700928
136 8.98984189033508
137 8.97649917602539
138 8.96181039810181
139 8.94367418289185
140 8.92970352172852
141 8.91647939682007
142 8.95909705162048
143 8.95546154975891
144 8.95323185920715
145 8.93683910369873
146 8.91739997863769
147 8.89754934310913
148 8.88061966896057
149 8.86554279327393
150 8.84501361846924
151 8.83433227539063
152 8.82492060661316
153 8.81752195358276
154 8.80928316116333
155 8.8015109539032
156 8.79397048950195
157 8.78659353256226
158 8.77897129058838
159 8.77216382026672
160 8.76544809341431
161 8.75828099250793
162 8.75543169975281
163 8.74632701873779
164 8.73793091773987
165 8.72998747825623
166 8.72202911376953
167 8.71491537094116
168 8.70841479301453
169 8.70249629020691
170 8.69704999923706
171 8.69193587303162
172 8.68703508377075
173 8.68226976394653
174 8.67759804725647
175 8.67300834655762
176 8.66849737167358
177 8.66414194107056
178 8.65989489555359
179 8.65576553344727
180 8.65156631469727
};
\addlegendentry{\texttt{BiSMO-CG} mean}
\addplot [thick, darkorange25512714]
table {%
60 16.4952732086182
61 15.7828498840332
62 14.9963419914246
63 14.5058230400085
64 13.8265824317932
65 13.1410750389099
66 12.6625178337097
67 12.247948551178
68 11.9250640869141
69 11.7376078605652
70 11.5616893768311
71 11.3492874145508
72 11.2063625812531
73 11.4241295337677
74 11.1854295253754
75 10.8759454727173
76 10.4993182182312
77 10.2960285186768
78 10.1424075126648
79 10.0277170658112
80 9.94220814704895
81 9.87646970748901
82 9.82613372802734
83 9.78463010787964
84 9.77655096054077
85 9.7286593914032
86 9.67623944282532
87 9.62795853614807
88 9.5811327457428
89 9.53649082183838
90 9.49497647285461
91 9.45411019325256
92 9.42176141738892
93 9.39198684692383
94 9.35502166748047
95 9.33016638755798
96 9.30076084136963
97 9.26682262420654
98 9.2384801864624
99 9.21326403617859
100 9.1901623249054
101 9.17106776237488
102 9.1529797077179
103 9.13675723075867
104 9.12144861221313
105 9.11363263130188
106 9.13876938819885
107 9.12201228141785
108 9.09758329391479
109 9.08972539901733
110 9.06879324913025
111 9.0445529460907
112 9.01816244125366
113 8.99970359802246
114 8.98645281791687
115 8.97536120414734
116 8.9648446559906
117 8.95410585403442
118 8.93954138755798
119 8.96740427017212
120 8.95463337898254
121 8.93658132553101
122 8.92390766143799
123 8.90429630279541
124 8.88893384933472
125 8.87605876922607
126 8.86502475738525
127 8.85546936988831
128 8.84685406684875
129 8.83757362365723
130 8.83033194541931
131 8.82299208641052
132 8.8147976398468
133 8.80694408416748
134 8.79923253059387
135 8.80972423553467
136 8.80150103569031
137 8.78278412818909
138 8.77281904220581
139 8.76446499824524
140 8.75736637115478
141 8.75167565345764
142 8.74669990539551
143 8.75178184509277
144 8.74589986801147
145 8.73979992866516
146 8.73353810310364
147 8.72730631828308
148 8.72139139175415
149 8.71596322059631
150 8.71112112998962
151 8.70681090354919
152 8.71281561851501
153 8.75718116760254
154 8.74030847549438
155 8.80210313796997
156 8.76627712249756
157 8.72696585655212
158 8.6861213684082
159 8.67781572341919
160 8.67290811538696
161 8.67506141662598
162 8.68191833496094
163 8.67433242797852
164 8.66880869865417
165 8.66221408843994
166 8.65595049858093
167 8.64917426109314
168 8.64282598495483
169 8.63660216331482
170 8.63044443130493
171 8.62424492835999
172 8.61807513237
173 8.61212687492371
174 8.60655827522278
175 8.60170178413391
176 8.59785828590393
177 8.59443941116333
178 8.59127688407898
179 8.58807125091553
180 8.6534074306488
};
\addlegendentry{\texttt{BiSMO-NMN} mean}
\end{axis}

\end{tikzpicture}

%% file: doc/conclu.tex
\section{Conclusion}
In this paper, we first establish a unified SMO framework utilizing Abbe-imaging, enabling simultaneous SO and MO gradient computation with enhanced lithographic precision and rapid calculation.
Building on this foundation, \texttt{BiSMO} is introduced, conceptualizing SMO as a bilevel problem and proposing three innovative methods for calculating source-mask best-response gradients, effectively addressing bilevel SMO challenges.
\texttt{BiSMO}'s gradient-based approach, with its global perspective and improved exploration of solution space
facilitates navigation out of local minima, ensuring better and faster converging SMO outcomes.
This method surpasses traditional \texttt{AM-SMO} limitations, positioning bilevel SMO as a future promising mainstream approach in the field.